\documentclass[twocolumn]{aastex631}

\newcommand\degree{^\circ}

\newcommand\xone{x_1}
\newcommand\xtwo{x_2}
\newcommand\xthree{x_3}

\newcommand\kms{\; {\rm km}\;{\rm s}^{-1}}
\newcommand\pc{\;{\rm pc}}
\newcommand\Mpc{\;{\rm Mpc}}

\newcommand\kpc{\;{\rm kpc}}

\newcommand\freq{\kms\kpc^{-1}}

\newcommand\C{{\mathcal C}}

\newcommand\simgt{\lower.5ex\hbox{$\; \buildrel > \over \sim \;$}}
\newcommand\simlt{\lower.5ex\hbox{$\; \buildrel < \over \sim \;$}}
\newcommand{\RNum}[1]{\uppercase\expandafter{\romannumeral #1\relax}}

\newcommand\tiltedRC{\mathrm{RC_{tilted}}}
\newcommand\trueRC{\mathrm{RC_{true}}}

\newcommand\dip{``dip''}
\newcommand\DIP{``DIP''}
\newcommand\misaell{``misaligned ellipses''}
\newcommand\aell{``aligned ellipses''}

\newcommand\MisaEllt{``MISALIGNED ELLIPSES''}

\newcommand\Misaellt{``Misaligned Ellipses''}
\newcommand\Aellt{``Aligned Ellipses''}

\newcommand\Diskfit{{\tt DiskFit}}

\shorttitle{Bar Effects on Galactic Rotation Curves}
\shortauthors{Liu et al.}
\graphicspath{{./}{figures/}}

\begin{document}

\title{The Impact of Bar-induced Non-Circular Motions on the Measurement of Galactic Rotation Curves}


\author[0000-0003-4052-8785]{Jie Liu}
\affiliation{Shanghai Astronomical Observatory, Chinese Academy of Sciences, 80 Nandan Road, Shanghai 200030, P.R. China}
\affiliation{School of Astronomy and Space Sciences, University of Chinese Academy of Sciences, 19A Yuquan Road, Beijing 100049, P.R. China}

\author[0000-0002-0627-8009]{Zhi Li}
\email{lizhi@shnu.edu.cn}
\affiliation{Shanghai Key Lab for Astrophysics, Shanghai Normal University, 100 Guilin Road, Shanghai 200234, People's Republic of China}
\affiliation{Department of Astronomy, School of Physics and Astronomy, Shanghai Jiao Tong University, 800 Dongchuan Road, Shanghai 200240, People's Republic of China}

\author[0000-0001-5604-1643]{Juntai Shen}
\email{jtshen@sjtu.edu.cn}
\affiliation{Department of Astronomy, School of Physics and Astronomy, Shanghai Jiao Tong University, 800 Dongchuan Road, Shanghai 200240, People's Republic of China}
\affiliation{Key Laboratory for Particle Astrophysics and Cosmology (MOE) / Shanghai Key Laboratory for Particle Physics and Cosmology, Shanghai 200240, People's Republic of China}

\received{October 31, 2024} \accepted{January 20, 2025}


\begin{abstract}
We study the impact of bar-induced non-circular motions on the derivation of galactic rotation curves (RCs) using hydrodynamic simulations and observational data from the PHANGS-ALMA survey. We confirm that non-circular motions induced by a bar can significantly bias RCs derived from the conventional tilted-ring method, consistent with previous findings. The shape of the derived RC depends on the position angle difference ($\Delta \phi$) between the major axes of the bar and the disk in the face-on plane. For $\left|\Delta \phi\right|\lesssim40\degree$, non-circular motions produce a bar-induced {\dip} feature (rise-drop-rise pattern) in the derived RC, which shows higher velocities near the nuclear ring and lower velocities in the bar region compared to the true RC ($\trueRC$). We demonstrate convincingly that such dip features are very common in the PHANGS-ALMA barred galaxies sample. Hydrodynamical simulations reveal that the {\dip} feature is caused by the perpendicular orientation of the gas flows in the nuclear ring and the bar; at low $\left|\Delta \phi\right|$ streamlines in the nuclear ring tend to enhance $V_\mathrm{los}$, while those in the bar tend to suppress $V_\mathrm{los}$. We use a simple {\misaell} model to qualitatively explain the general trend of RCs from the tilted-ring method ($\tiltedRC$) and the discrepancy between $\tiltedRC$ and $\trueRC$. Furthermore, we propose a straightforward method to implement a first-order correction to the RC derived from the tilted-ring method. Our study is the first to systematically discuss the bar-induced {\dip} feature in the RCs of barred galaxies combining both simulations and observations.
\end{abstract}

\section{Introduction} \label{sec:intro}

Galactic circular rotation curves (RCs) are crucial for understanding galactic properties. They provide the first clear direct evidence for the existence of dark matter by suggesting that galaxies contain more mass than visible matter alone (\citealt{1978PhDT.......195B}). Various methods have been developed to derive accurate RCs from galactic kinematics, such as the centroid-velocity method (\citealt{1982ApJ...261..439R}), the peak-intensity-velocity method (\citealt{1992AJ....103.1552M}), the terminal-velocity and envelope-tracing method (\citealt{2017PASJ...69R...1S}), the iteration method (\citealt{2000ApJ...534..670T}), and the tilted-ring method (\citealt{1974ApJ...193..309R,1992ASPC...25..131V}). These methods are commonly applied to derive RCs from the kinematics of gas disks, which are assumed to have lower velocity dispersion and more circular orbits than the stellar component.

However, when a galaxy has a rotating stellar bar, non-circular motions become significant around the bar region. This is commonly seen in both hydrodynamic simulations (\citealt{1980ApJ...235..803S,1992MNRAS.259..345A,1997MNRAS.292..349S,2004MNRAS.354..883M,2004MNRAS.354..892M,2012ApJ...758...14K,2012ApJ...747...60K,2015ApJ...806..150L}) and observations (e.g., H\,{\footnotesize I} in \citealt{2008AJ....136.2563W}, and CO in \citealt{2021ApJS..257...43L}). In this case, the circular rotation curve is hard to define and may not exactly reflect the real mass distribution as the potential is non-axisymmetric. In addition, large non-circular motions in the gas disk may violate the underlying assumption of the aforementioned methods, leaving uncertainties in the derived RC. It is important to note that the galactic properties directly inferred from RCs may be subject to these uncertainties; for example, the orbital time, $t_{orb}$, in the Molecular Elmegreen–Silk Relation \citep{1997RMxAC...6..165E,1997ApJ...481..703S,2023ApJ...945L..19S}, and the dark matter core size in dwarf galaxies \citep{2019MNRAS.482..821O}. Properly addressing these deviations is crucial to obtaining a more accurate rotation curve as well as other galactic parameters. 

On the other hand, studies aimed at quantitatively correcting the effects of non-circular motions on RCs are incomplete.  \citet{2004ApJ...617.1059R} conducted $N$-body simulations for barred galaxies and derived RCs using the tilted-ring method. They found that the method could cause an $\sim20\%$ underestimation of the rotation curve in the central region. \citet{2015A&A...578A..14C} studied the RC of the Milky Way and found that the deviation of the RC is related to the ``viewing angle'' of the bar, aligning with the previous results by \citet{1991MNRAS.252..210B} and \citet{1997MNRAS.292..349S}. They proposed a straightforward method to correct the RC with a radially dependent factor $\left(\frac{v(R)}{v_\mathrm{true}}\right)_\mathrm{sim}$ derived from simulations. \citet{2015arXiv150907120S} developed {\Diskfit} to extract non-circular motions with an $m=2$ (i.e., bisymmetric) component. \citet{2015MNRAS.454.3743R,2016A&A...594A..86R} tested this tool with simulations and concluded that {\Diskfit} could roughly recover the true RC when the bar is not aligned or perpendicular to the disk major axis, although their simulation cannot fully resolve the gas substructures generated by the bar\footnote{We conducted tests using {\Diskfit} and further discuss the limitations of \citet{2015MNRAS.454.3743R,2016A&A...594A..86R} in Appendix~\ref{subsec:DiskFit}.}. Nevertheless, a more comprehensive study of the RCs in barred galaxies, along with the necessary corrections, is still needed.

This paper aims to deepen the understanding of the systematic deviation of derived RCs of barred galaxies based on both hydrodynamical simulations and observations. We define the RC of a barred galaxy as the circular velocity in centrifugal balance with the azimuthally averaged gravitational potential \citep[see the discussions in][]{2007ApJ...664..204S}. We further propose a straightforward correction method for RCs derived from the tilted-ring method in barred galaxies. The paper is organized as follows: \S\ref{sec:noncirsim} presents an analysis of the non-circular motion effects based on simulations. \S\ref{sec:noncirobs} demonstrates the bar-induced {\dip} features in the observed RCs of real galaxies. \S\ref{sec:discuss} introduces two simple models to illustrate the physical origin of the RC features and proposes a straightforward correction method. The summary is in \S\ref{sec:summary}. 

\section{The effect of Non-circular motions in Simulations}\label{sec:noncirsim}
\begin{figure*}[ht]
    \centering
    \includegraphics[width=1.0\textwidth]{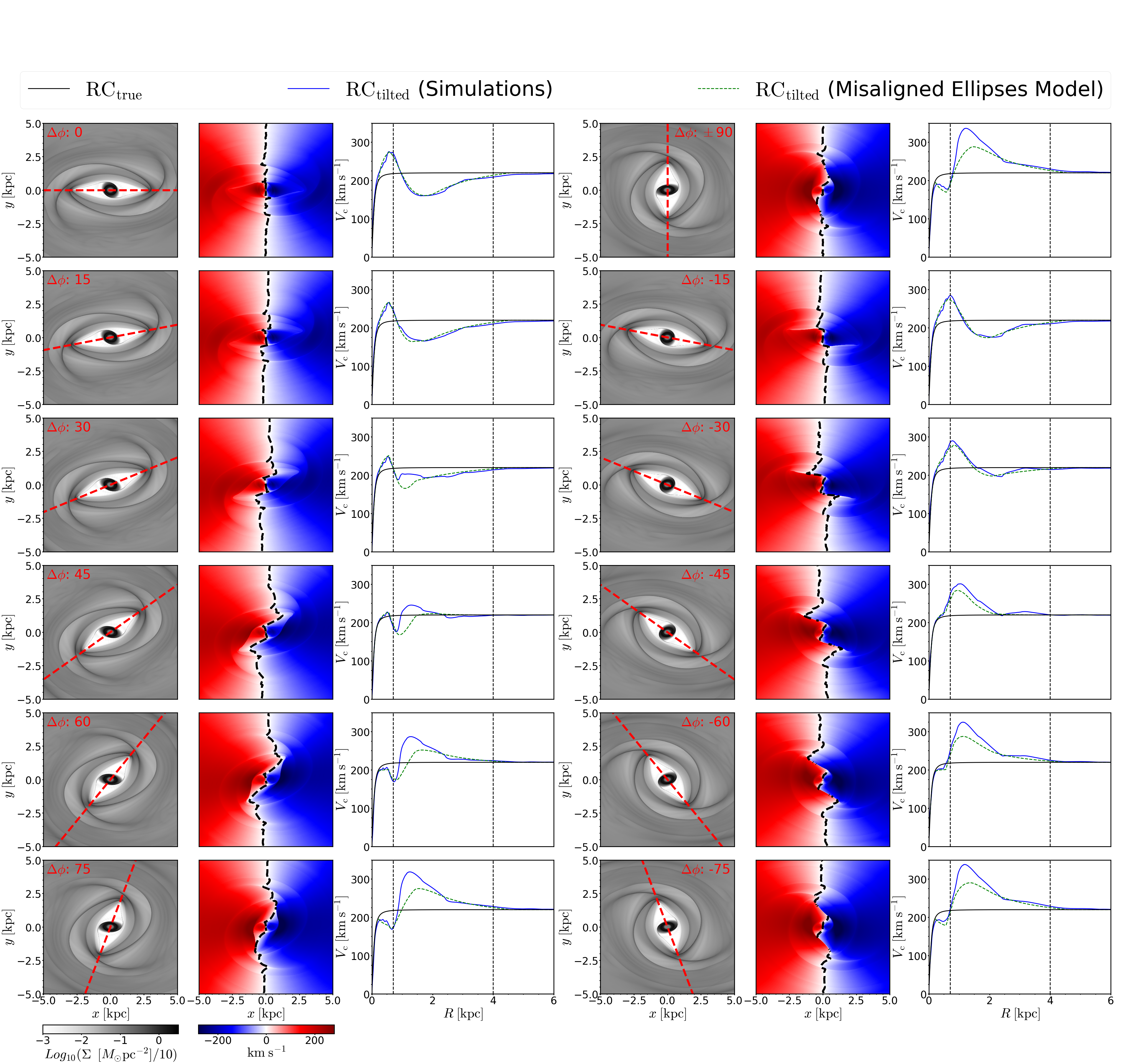}
    \caption{Clockwise-rotating quadrupole models. Columns 1 and 4 display the projected surface density, with bar directions indicated by red dashed lines. $\Delta \phi$ from $0\degree$ to $\pm\; 90\degree$ is indicated in the top corner of each plot. Columns 2 and 5 display the constructed $V_{\mathrm{los}}$ map generated directly from the components $V_x$ and $V_y$ with an inclination (\textit{i}) of $45\degree$. The zero-velocity lines are represented by black dashed lines. Columns 3 and 6 present a comparison between $\trueRC$ and $\tiltedRC$ (derived from $V_{\mathrm{los}}$ using the tilted-ring method). $\tiltedRC$ derived from simulation is represented by the blue solid line, while the other version (the green dashed line) is derived from the {\misaell} model (see \S~\ref{subsec:misalign_ellipse}). The inner and outer vertical black dashed lines denote the nuclear ring size and the bar length, respectively.}
    \label{fig:Sormani_Ferrers_RC}
\end{figure*}
We first illustrate the effects of bar-induced non-circular motions on rotation curves using simple hydrodynamical simulations. We test two classical barred galactic potentials: the quadrupole model and the Ferrers model.

\textit{The quadrupole model} (\citealt{1991MNRAS.252..210B,2015MNRAS.454.1818S}) consists of the quadrupole term: 
\begin{equation}
\rho_2(r,\theta,\phi)=\frac{v_0^2A}{4\pi G r_q^2}\exp{\left(2-\frac{2r}{r_q}\right)}\sin^2{\theta}\cos{2\phi},
\end{equation}
 and the monopole term:
\begin{equation}
\rho_0(x,y)= \frac{v_0^2}{2\pi G}\frac{R_e^2}{(R_e^2+x^2+y^2)^2},
\end{equation}
where $A=0.6,\ r_q=2.0\kpc,\ R_e=0.1\kpc$, and $v_0=220 \kms$.

We conduct hydrodynamical simulations of a 2D rotating isothermal gas disk in the above potential. The simulations are performed with the grid-based magnetohydrodynamics (MHD) code, {\tt Athena++} \citep{2020ApJS..249....4S}, on a uniform Cartesian grid. The pattern speed ($\Omega_{\mathrm{b}}$) of the quadrupole model is $40\freq$, giving the co-rotation radius $R_{\rm CR}= 5.5\kpc$. The spatial resolution is $9.76\pc$, and the effective sound speed is $10\kms$. The bar grows over three bar rotation periods ($3\times2\pi/\Omega_b$). Assuming a fixed inclination ($i=45\degree$), we construct a $V_{\rm los}$ map using the snapshot at $t=1$ Gyr when a quasi-steady gas flow pattern has been well developed \footnote{$V_{\rm los}$ is generated directly from the components $V_x$ and $V_y$, as for 2D simulations we do not have multiple density and velocity values for each pixel.}.

We apply \textit{the tilted-ring method} by fitting the $V_{\mathrm{los}}$ map in polar coordinates with a commonly used equation: 
\begin{equation}
    V_{\mathrm{los}}(R,\phi)=V_{\mathrm{sys}}+V_{\mathrm{c}}(R)\cos{\phi}\sin{i},
    \label{eq:RC_fit}
\end{equation}
where $V_{\mathrm{sys}}=0\kms,\ i=45\degree,\ V_{\mathrm{c}}(R)$ are the systemic velocity, the inclination angle, and the circular velocity profile of the galaxy, respectively. The position angle (PA) of the disk major axis is set at $\phi=0\degree$, thus fixing $\mathrm{PA_{disk}}=0\degree$. We define $\Delta \phi$ as the angle between the major axes of the bar and the disk of the galaxy in the face-on plane, and vary $\Delta \phi$ from $-90\degree$ to $90\degree$. Mirror symmetry ensures that the gas flow pattern produces identical effects in counter-clockwise rotating cases with a positive $\Delta \phi$ and in clockwise cases with a negative $\Delta \phi$. For clarity, only the clockwise rotating quadrupole model is shown in Figure~\ref{fig:Sormani_Ferrers_RC}.

Note that $\Delta \phi$ is defined in the face-on (deprojected) view, unless otherwise specified. $\Delta \phi$ is related to the angle difference ($\Delta \mathrm{PA}$) in the sky (or projected) plane by the following equations assuming a planar 2D disk: 
\begin{equation}
\tan{\Delta \phi}=\tan{\Delta \mathrm{PA}}/\cos{i},\; \Delta \mathrm{PA}=\mathrm{PA_{bar}-PA_{disk}}.
\label{eq:PApro_face}
\end{equation}

As shown in Figure~\ref{fig:Sormani_Ferrers_RC}, different $\rm{\Delta \phi}$ lead to different orientations of the gas flow pattern. The features in the $\rm{V_{los}}$ map also change accordingly with different twisted zero-velocity lines. The RC from the tilted-ring method, $\tiltedRC$ (blue solid line), which assumes circular motions, clearly deviates from $\rm{RC_{true}}$. In the very central region ($R<250\pc$) $\tiltedRC$ aligns closely with $\trueRC$. At larger radii, the deviation can be as large as $\sim100\kms$ and the exact value depends strongly on $\Delta \phi$. Within the bar region $\tiltedRC$ is considerably lower than $\trueRC$ when the bar major axis aligns closely with the disk major axis. Conversely, $\tiltedRC$ becomes significantly higher than $\trueRC$ when the bar major axis is perpendicular to the disk major axis. In general, $\tiltedRC$ rises rapidly to its peak in the center, then declines to a local minimum and eventually rises again matching $\rm{RC_{true}}$ at $R>\sim4\kpc$. We define the central peak and the following decline-rise trend as a {\dip} feature in the RC (Figure~\ref{fig:Sormani_Ferrers_RC}), which is more prominent when $\left|\Delta \phi\right|\lesssim 40\degree$. Outside this $\phi$ range, the {\dip} feature becomes less obvious and more like a single high peak. It is important to note that the definition of the {\dip} feature not only relies on a decline-rise trend in the RC, but also requires that the central peak is higher than $\trueRC$ and the local minimum is lower than $\trueRC$. The amplitude of the velocity peak for the negative $\Delta \phi$ is generally higher compared to the positive $\Delta \phi$, which will be discussed in more detail in \S~\ref{subsec:misalign_ellipse}. The shape of $\tiltedRC$ depends sensitively on $\Delta \phi$. These findings align with the conclusions drawn by \citet{2004ApJ...617.1059R}, \citet{2015A&A...578A..14C}, and \citet{2015MNRAS.454.3743R,2016A&A...594A..86R}. 

\begin{figure}
    \centering
    \includegraphics[width=1.0\columnwidth]{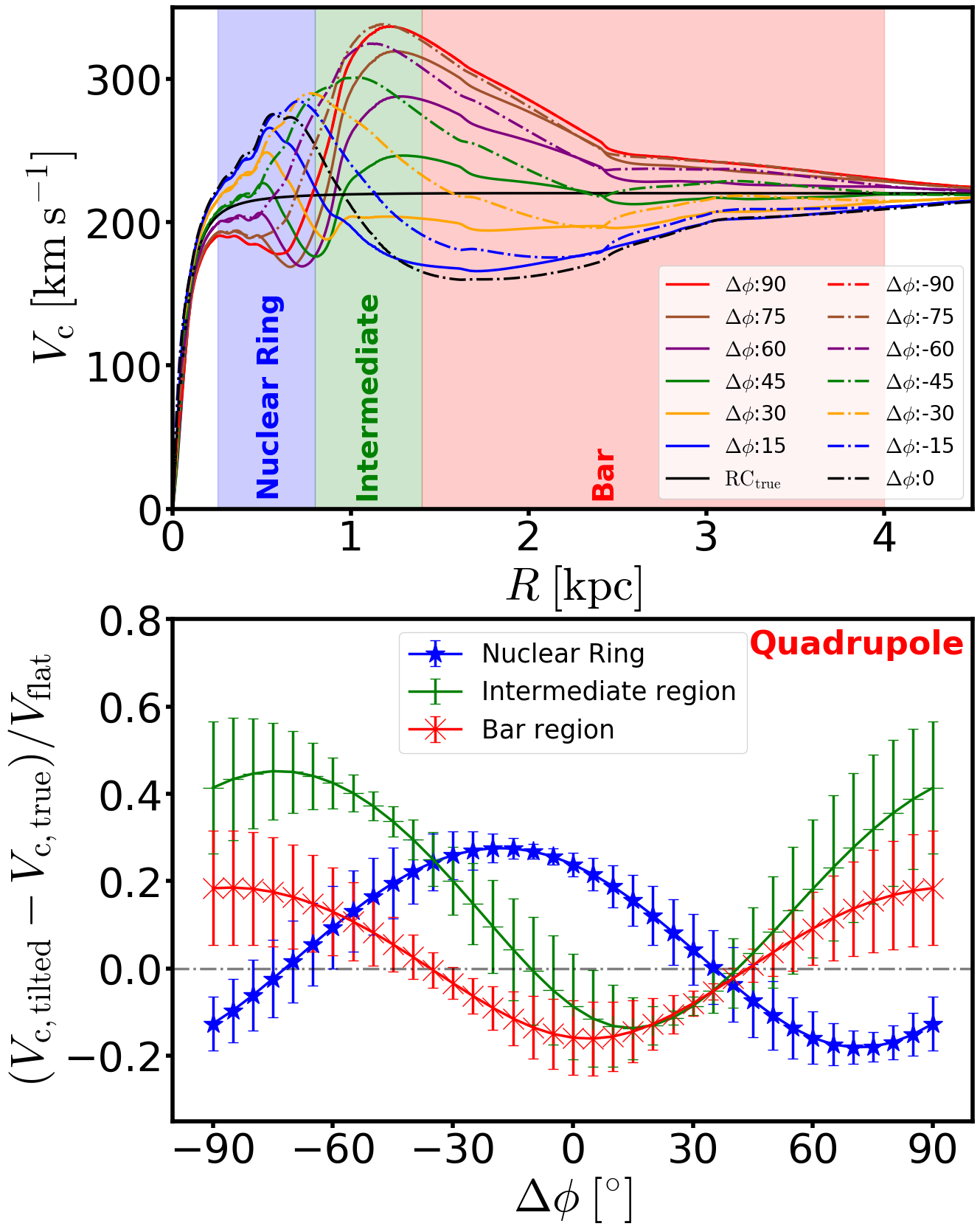}
    \caption{Deviation of RCs for the quadrupole model. The top panel illustrates a comparison between $\trueRC$ and $\tiltedRC$ across various $\Delta \phi$ cases, each represented by a different color. The shaded regions in blue, green, and red represent the nuclear ring, the intermediate, and the bar region, respectively. In the bottom panel, the $x$-axis represents $\Delta \phi$ while the $y$-axis represents the deviation of the RC, calculated as $(V_{\mathrm{c,tilted}}-V_{\mathrm{c,true}})/V_{\mathrm{flat}}$, where $V_{\mathrm{c,tilted}}$ and $V_{\mathrm{c,true}}$ stand for the values in $\tiltedRC$ and $\trueRC$, and $V_{\mathrm{flat}} = 220 \kms$. The error bars represent the statistical standard deviation in each respective region. Small error bars indicate that the $\tiltedRC$ in the corresponding region deviates systematically from the $\rm{RC_{true}}$.}
    \label{fig:RC_deviation}
\end{figure}

A more quantitative analysis of the deviation of RC in the quadrupole model is shown in Figure~\ref{fig:RC_deviation}. We divide the region within the bar length into three zones: the nuclear ring $(0.25-0.8\kpc)$, the intermediate $(0.8-1.4\kpc)$, and the bar $(1.4-4.0\kpc)$ regions. In the quadrupole model, the bar length is not explicitly defined; we designate the empirical bar boundary as $2r_q=4.0\kpc$. In the bottom panel, we calculate the average deviation (defined as $(\tiltedRC-\trueRC)/v_0$) in each region and investigate its relationship to $\Delta \phi$. Figure~\ref{fig:RC_deviation} clearly suggests that the {\dip} feature (where the nuclear ring region has higher $\tiltedRC$, and the intermediate and bar regions have lower $\tiltedRC$ compared to $\trueRC$) only appears when $\Delta\phi$ ranges from $-40\degree$ to $40\degree$. This range does not represent a strict limit but provides an approximate boundary for the {\dip} feature to be present. This range depends on the strength and the pattern speed of the bar as well.

The three lines in the bottom panel of Figure~\ref{fig:RC_deviation} show different deviation trends in the three regions. Specifically, the deviation profile of the nuclear ring region presents a pattern nearly opposite to that observed in the bar region. The profile of the intermediate region is similar to the bar region but shows a stronger asymmetry and a larger amplitude. This discrepancy results from the different patterns of non-circular motions in the three regions. The elliptical nuclear ring is almost perpendicular to the bar \citep{1992MNRAS.259..328A,1992MNRAS.259..345A,2004MNRAS.354..883M,2004MNRAS.354..892M,2012ApJ...758...14K,2012ApJ...747...60K,2015ApJ...806..150L}. Therefore, when the bar is parallel to the disk major axis ($\Delta\phi\sim0\degree$), the streaming motions contribute more to $V_{\rm los}$ in the nuclear ring region but less in the bar region, resulting in a central peak followed by a decline-rise trend. When the bar is perpendicular to the disk major axis ($\Delta\phi\sim90\degree$), the contribution of streaming motions reverses in the two regions, leading to a lower value in the nuclear region followed by a high peak. We note that the deviation of the RC across the three regions shows an asymmetry with respect to $\Delta\phi=0\degree$. These trends and the asymmetry can be understood by a simplified misaligned ellipses model discussed in \S~\ref{subsec:misalign_ellipse}.

We also tested the \textit{Ferrers model}, which consists of a Ferrers bar (\citealt{1984A&A...134..373P}), a Miyamoto-Nagai disk \citep{1975PASJ...27..533M}, and a modified-Hubble-profile bulge. This model has been well studied previously \citep[e.g.,][]{1992MNRAS.259..328A,1992MNRAS.259..345A,2004MNRAS.354..883M,2004MNRAS.354..892M,2012ApJ...758...14K,2012ApJ...747...60K,2015ApJ...806..150L}. For brevity this paper focuses on the results of the quadrupole model, as the Ferrers model yields similar effects of non-circular motions on RCs.

To further validate our findings we conduct additional tests using simulation data. Observations can only detect signals with sufficient signal-to-noise (SN) ratios due to the sensitivity limitations of telescopes. Consequently, the observed $V_{\rm los}$ map often contains low-SN pixels represented as {\tt NaN} values. To mimic this effect, we apply a simple density threshold in our simulations. Specifically, we assign $V_{\rm los}$ of those pixels with densities below a critical value to {\tt NaN}. This approach allows us to generate mock images similar to real observational data. The results of these tests are presented in Appendix~\ref{Appendix:dens_threshold}. The {\dip} feature is still present and becomes even more prominent with a density threshold cut.

To sum up, the simulation results suggest a {\dip} feature in the RC of a barred galaxy under specific $\Delta \phi$ configurations. A natural question to pursue next is: Can similar bar-induced {\dip} features also be found in observations?

\section{The bar-induced {\DIP} features in observations} 
\label{sec:noncirobs}

\begin{figure*}
    \centering
    \includegraphics[width=1.0\textwidth]{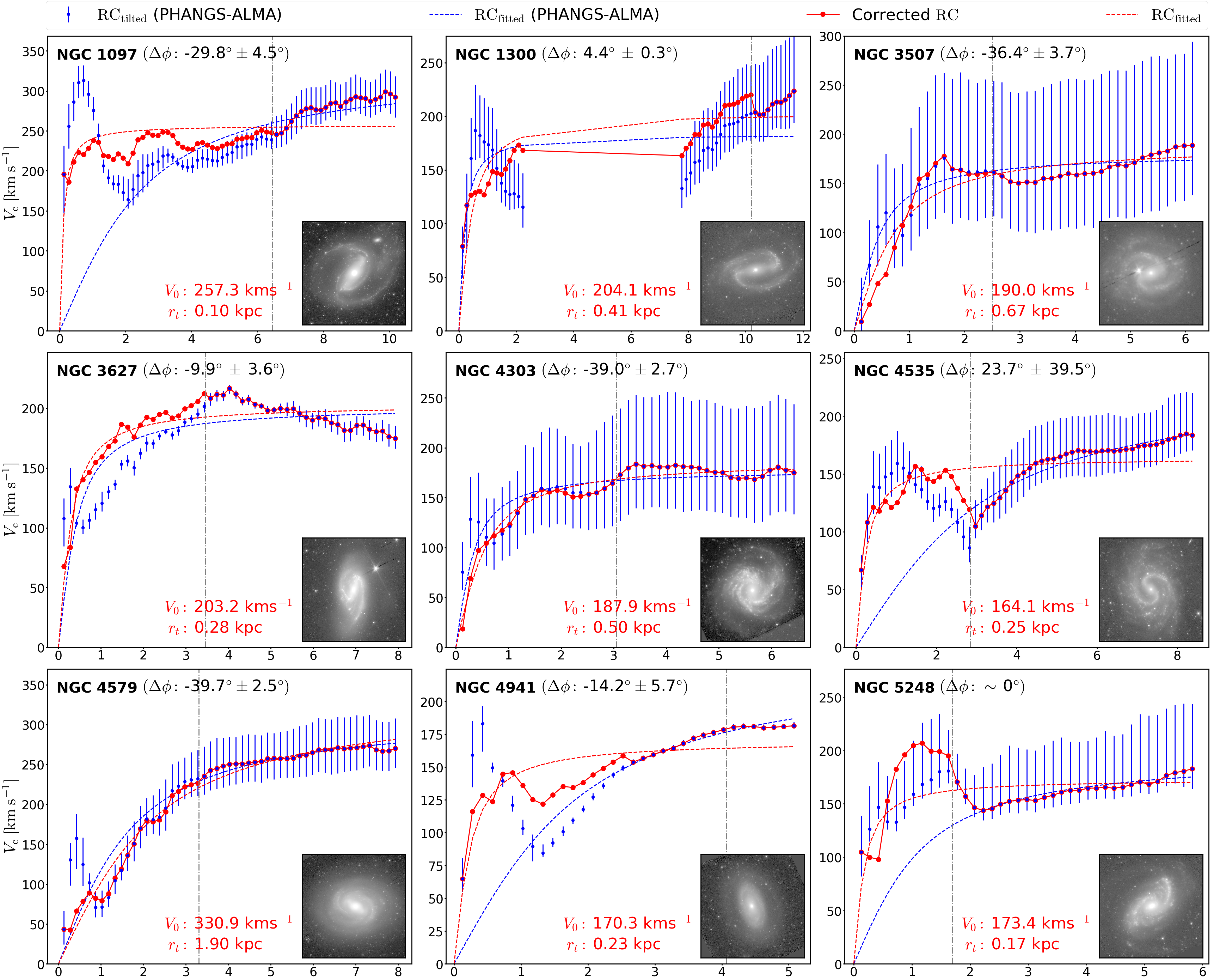}
    \caption{RCs for real galaxies with a bar-induced {\dip} feature. The blue dots with error bars represent $\tiltedRC$ from \cite{2020ApJ...897..122L}, while the blue dashed lines illustrate their best fit of Eq.~\ref{eq:RCfit}. The red dots are our corrected RCs with the correction procedure described in \S~\ref{subsec:RCcorrection}, and fitted using the same equation shown by the red dashed lines. The fitted $V_0$ and $r_\mathrm{t}$ for the red dots are shown in the bottom of each panel. The vertical grey dot-dashed lines indicate the bar length of each galaxy from \cite{2021A&A...656A.133Q} or our re-determination. We miss the nuclear ring size as its measurement in real galaxies is challenging due to the insufficient spatial resolution in observations. The value of $\Delta \phi$ (deprojected by Eq.~\ref{eq:PApro_face}) and the S$^4$G $3.6\mu \mathrm{m}$ image for each galaxy are shown at the bottom of each subplot.}
    \label{fig:dip_galaxies}
\end{figure*}

The detection of the {\dip} features in rotation curves requires data with sufficient spatial resolution to resolve the non-circular motions in the bar region. This paper utilizes the high-resolution ($\sim0.5^{\prime \prime}$ or $\sim240\pc$ for a galaxy at a distance of $10\Mpc$) RCs of 67 spiral galaxies from the PHANGS-ALMA survey (\citealt{2021ApJS..257...43L,2020ApJ...897..122L}). The inclination and PA of the bar and disk for most of these galaxies have been determined by \cite{2015ApJS..219....4S} using ellipse fitting on S$^4$G $3.6\mathrm{\mu m}$ photometric images \citep{2010PASP..122.1397S}.

We first adopt the morphology classification and bar length of these galaxies from \cite{2021A&A...656A.133Q}. However, we note several issues. (1) The bar length may not be accurate. The bar lengths of NGC~4321 and NGC~1433 are significantly underestimated due to their double-barred structure. The bar length value from \cite{2021A&A...656A.133Q} might indicate the length of the inner bar rather than the primary bar. We re-determine the lengths of the primary bars of NGC~4321 and NGC~1433 to be $4.9\kpc$ and $5.9\kpc$, respectively. These values are consistent with the measurements reported by \cite{2004A&A...415..941E} ($4.3\kpc$ and $6.0\kpc$). (2) Some galaxies have been incorrectly classified as either barred or unbarred. NGC~4941 and NGC~5248 which exhibit typical bar-driven gas flow patterns in the CO~(2-1) map are misclassified as unbarred galaxies. Consequently, we re-examine the S$^4$G photometric images using ellipse fitting and the CO images from the PHANGS-ALMA survey to determine the bar length and $\Delta \mathrm{PA}$. To compare with simulations we convert $\Delta \mathrm{PA}$ to $\Delta \phi$ using Eq.~\ref{eq:PApro_face}. The bar lengths of NGC~4941 and NGC~5248 are remeasured to be $3.6\kpc$ and $1.2\kpc$, respectively. On the other hand, NGC~1559 and NGC~4654 may not host a strong bar, as evidenced by the absence of dust lanes and nuclear rings in their gas flow patterns, but further observations are required to validate this assertion. In summary, we update the bar lengths of NGC~4321, NGC~1433, NGC~4941, NGC~5248, and re-classify NGC~4941 (barred), NGC~5248 (barred), NGC~1559 (unbarred), NGC~4654 (unbarred).

We then exclude 6 edge-on ($i\geq 70\degree$) galaxies (NGC~1511, NGC~1546, NGC~3137, NGC~3511, NGC~4569, NGC~4951) due to large uncertainties in $\Delta\phi$ measurement, and 5 face-on ($i\leq 20 \degree$) galaxies (NGC~0628, NGC~1672, NGC~3059, NGC~4457, NGC~4540) due to large uncertainties in their RCs from our sample. Additionally, we exclude 11 galaxies where little CO is present in the bar region. Based on earlier discussions about the selection and re-evaluation of bar lengths and classifications, our study comprises a sample of 45 galaxies, including 29 barred and 16 unbarred galaxies.

The identification of a typical bar-induced {\dip} feature in real galaxies is limited by the unknown $\trueRC$. This feature is recognized primarily based on the decline-rise trend in the $\tiltedRC$. According to the results in \S~\ref{sec:noncirsim} and our other simulations with different bar strength and pattern speed, we empirically use $\left|\Delta \phi\right|\lesssim40\degree$ as the optimal $\Delta \phi$ range for a {\dip} feature considering various bar properties in real galaxies.

\subsection{Barred galaxies with a {\dip} feature}
Based on the criterion of $\left|\Delta \phi\right|\lesssim40\degree$, typical bar-induced {\dip} features in the RCs of 9 barred galaxies (NGC~1097, NGC~1300, NGC~3507, NGC~3627, NGC~4303, NGC~4535, NGC~4579, NGC~4941 and NGC~5248) have been identified, as shown in Figure~\ref{fig:dip_galaxies}. The 
{\dip} feature is strong in NGC~1097, NGC~1300, and NGC~4941, indicating a robust bar in these galaxies, while it is less pronounced in NGC~3507, NGC~3627, and NGC~4303, implying a weak bar. The decline beyond the bar length in NGC~5248 may be associated with the spirals outside its bar. Apart from the 9 barred galaxies, the three outliers NGC~4321, NGC~1566, and NGC~2997 suggest different mechanisms may generate a decline-rise feature in the RCs.

\begin{figure*}
    \centering
    \includegraphics[width=1.0\textwidth]{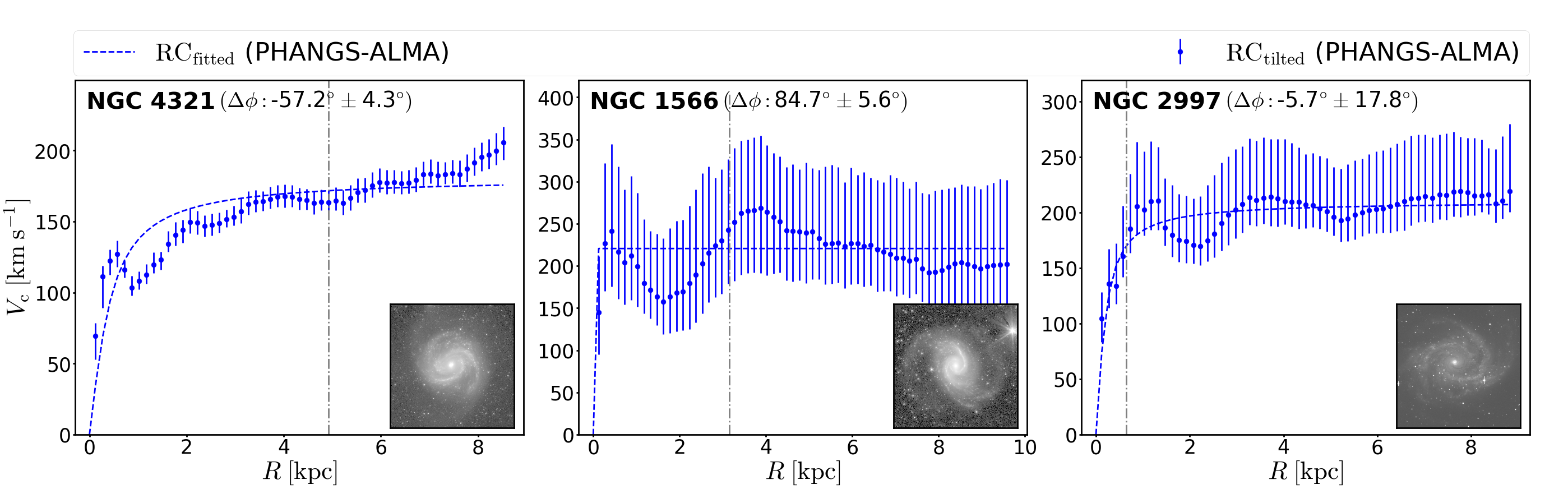}
    \caption{RCs for outlier galaxies with a decline-rise feature. The legend details are similar to those in Figure~\ref{fig:dip_galaxies}, but the RC correction is not applicable to these galaxies.}
    \label{fig:galaxies_wpa}
\end{figure*}

NGC~4321 and NGC~1566 exhibit a decline-rise feature as shown in Figure~\ref{fig:galaxies_wpa}, but their associated $\Delta \phi$ values significantly exceed the optimal $\Delta \phi$ range for a {\dip} feature. According to \citet{2015ApJS..219....4S}, the surface brightness decomposition reveals that NGC~4321 and NGC~1566 both possess a compact central component (effective radius $R_e\sim500\pc$ and $R_e\sim200\pc$) contributing $10\%$ and $5\%$ of the total flux, respectively. This central component could potentially result in a high peak in the central region of the RC, followed by a decline-rise feature in the RC. However, this feature is distinct from the {\dip} feature we previously defined, as the values in this feature may not necessarily be smaller than $\trueRC$. Additionally, the location of the feature is closely associated with the scale length of the bulge, and the central peak may rise more steeply compared to the bar-induced {\dip}. While bar-induced non-circular motions can generate {\dip} features, a massive bulge in the central region of a barred or unbarred galaxy may also result in similar features. These two galaxies highlight the challenge in identifying the {\dip} feature in the RC of real galaxies due to the unknown $\trueRC$. However, this bulge-induced feature is not observed in our unbarred sample, partially due to the small sample size (\S~\ref{subsec:unbarredgal}).

Regarding NGC~2997, the decline-rise feature appears outside the bar, potentially originating from non-circular motions influenced by the strong grand-design spiral structure, which has an inner Lindblad resonance of $\sim2.9\kpc$ \citep{2009A&A...499L..21G}. This decline-rise feature is evidently different from the {\dip} as it occurs outside the bar. Interestingly, NGC~2997 is the only one in our sample that has this RC feature possibly related to the spirals. We speculate that not all spirals can create such a decline-rise trend in the RC, as this may depend on the pitch angle as well as the strength. This serves as a cautionary note that the deviation of the RC due to the strong spirals cannot be completely overlooked. Complex spiral patterns could introduce additional complexities to the shapes of RCs, which are not addressed in this study. 

\subsection{Barred galaxies without a {\dip} feature} 

The RCs of barred galaxies without a {\dip} feature are presented in Appendix~\ref{sec:appendixb}. Most barred galaxies lacking the {\dip} feature have $\left|\Delta \phi\right|>40\degree$ (Table~\ref{tab:galaxy_no_dip}), except for IC~5273 and NGC~2903. These two galaxies have appropriate $\left|\Delta \phi\right|<40\degree$, but the {\dip} feature is not observed in their RCs. A possible explanation may be that the short bar length of these two galaxies (IC~5273: $1.4\kpc$, NGC~2903: $2.6\kpc$) may possess a very small nuclear ring ($r_\mathrm{ring}\lesssim 200\pc$) and require observations with higher spatial resolution ($\lesssim 50\pc$) to resolve the central region. 
\begin{table}[t]
    \begin{tabular}{cc|cc}
        \hline
        Galaxy &  $\Delta \phi$ & Galaxy & $\Delta \phi$\\
        \hline
        IC~1954 & $45.8\degree$ & NGC~1087 & $-56.6\degree$  \\
        NGC~1317 & $-84.9\degree$ & NGC~1365 & $-67.2\degree$  \\
        NGC~1433 & $-73.5\degree$ & NGC~2566 & $74.9\degree$  \\
        NGC~3351 & $-79.4\degree$ & NGC~4536 & $61.0\degree$  \\
        NGC~4548 & $88.2\degree$ & NGC~4781 & $-42.2\degree$  \\
        NGC~5042 & $46.9\degree$ & NGC~5134 & $76.7\degree$  \\
        NGC~5643 & $-43.0\degree$ & NGC~6300 & $67.9\degree$  \\
        NGC~7496 & $69.5\degree$ & \textcolor{red}{IC~5273} &  \textcolor{red}{$-16.5\degree$}\\
        \textcolor{red}{NGC~2903} & \textcolor{red}{$19.9\degree$} &  & \\ 
        \hline
    \end{tabular}
    \caption{Barred galaxies without a {\dip} feature and their $\Delta \phi$. The red-marked galaxies are the outliers with appropriate $\Delta \phi$ without {\dip}.}
    \label{tab:galaxy_no_dip}
\end{table}
\subsection{Unbarred Galaxies}\label{subsec:unbarredgal}
As expected, the {\dip} feature is not observed in any of the unbarred galaxies in our sample. The details of the RCs for these galaxies are shown in Appendix~\ref{sec:appendixb}.

\hspace{\fill}

Drawing upon the findings of our analysis, we can indeed identify the bar-induced {\dip} features in the RCs of real barred galaxies. About $85\%$ of the galaxies in our sample align with the expectations from simulations. We also find that the RC of a barred galaxy is affected not only by the bar-induced non-circular motions but also by the presence of a massive bulge inside the bar and/or strong spiral arms outside the bar. In the next section, we present a simple model to better elaborate on the origin of the bar-induced {\dip} feature and the deviation of the RCs.

\section{Discussion}\label{sec:discuss}
\subsection{{\Aellt} Model}
\label{subsec:aligned_ellipse}

\begin{figure}
    \centering
    \includegraphics[width=1.0\columnwidth]{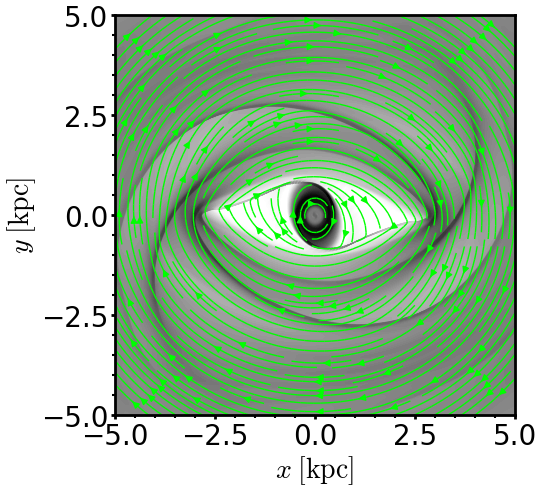}
    \caption{Gas streamlines (green) superimposed on the surface density of the quadrupole model. The bar major axis is along the $x$-axis ($R_{\rm CR}=5.5\kpc$).}
    \label{fig:streamlines} 
\end{figure}

The strong correlation between the deviation of the RC and the geometric parameter $\Delta \phi$ motivates a deeper examination of the characteristics of non-circular motions. Figure~\ref{fig:streamlines} shows the streamlines superimposed on the gas surface density of the quadrupole model. These streamlines can be approximated as a series of concentric ellipses, which vary in both axial ratio ($a/b$) and orientation at different radii. Near the bar end the axial ratio is larger, and the orientation aligns parallel to the bar major axis. In contrast, in the nuclear ring region the ellipses are more circular and perpendicular to the bar major axis. In the region between the nuclear ring and the bar end, the major axis orientations of these elliptical streamlines rotate counter-clockwise, shifting to be aligned with the bar major axis as the radius increases. The axial ratio of the streamlines also increases simultaneously. 

\begin{figure}
    \centering
    \includegraphics[width=0.9\columnwidth]{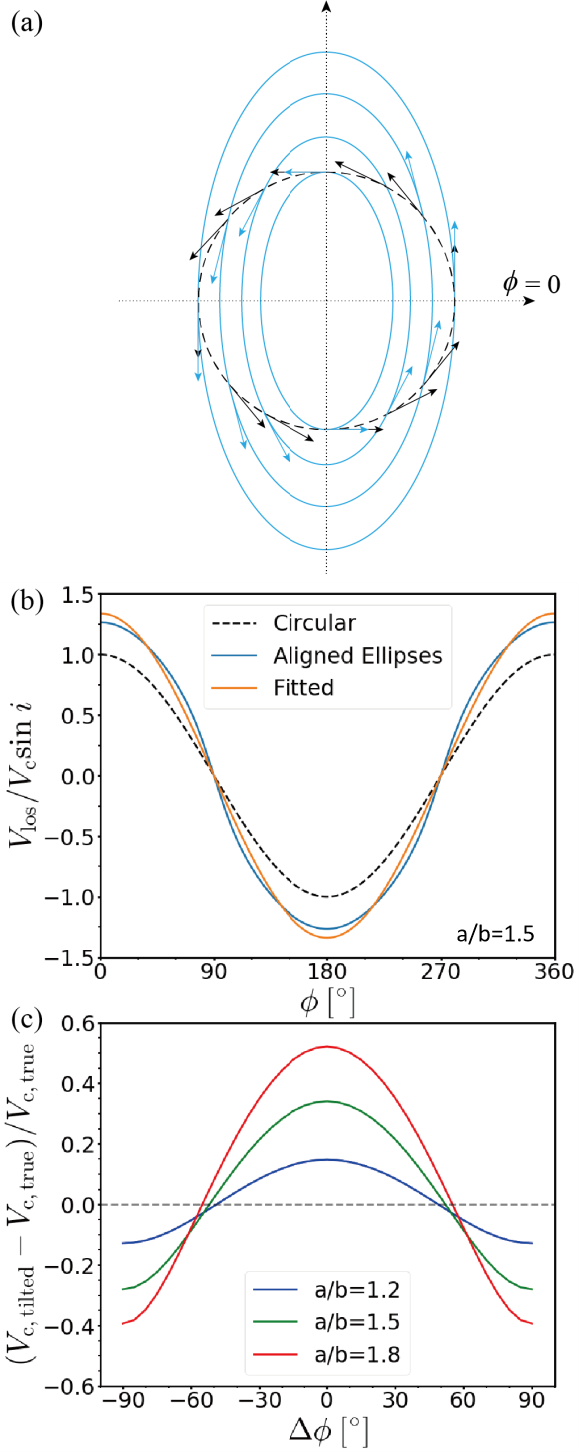}
    \caption{(a) A schematic diagram showing the difference in velocity between elliptical and circular orbits. (b) $V_{\rm los}$ at a specific radius ($R=0.8\kpc$) as a function of the azimuthal angle, obtained after projecting the velocity of circular (black dashed line) and elliptical (blue solid line) orbits. The orange solid line represents the fitted $V_{\rm los}$ using Eq.~\ref{eq:RC_fit}. (c) Deviation of rotation velocity in models of elliptical orbits with varying axial ratios, plotted against $\Delta \phi$.}
    \label{fig:explanation}
\end{figure}

Motivated by the configuration of the streamlines, we adopt a simple geometric model to explain the deviation of the RCs. We first aim to explain the deviation within the elliptical nuclear ring, where the streamline shapes are relatively simple. We begin with an {\aell} model, wherein all gas follows elliptical orbits with constant angular momentum (as illustrated in Figure~\ref{fig:explanation}a). In this model, we assume that the angular momentum of each orbit remains constant, calculated as the product of the average orbital radius and the circular velocity (denoted as $V_\mathrm{c}$) at that radius. The circular velocity profile is from the model represented by the black lines in Figure~\ref{fig:RC_deviation}. These elliptical orbits share the same axis ratio ($a/b=1.8$). Instead of being a single circular orbit at a certain radius, the velocity on the black dashed circle (at $R=0.8\kpc$) is determined by a series of elliptical orbits. The major axes of these elliptical orbits range from $0.8\kpc$ to $1.44\kpc$. The velocity directions of circular orbits and elliptical orbits are very different as shown by the black and blue arrows. In Figure~\ref{fig:explanation}(b), we project the velocity on the $R=0.8\kpc$ circle to obtain the $V_{\mathrm{los}}$ profile (the blue solid line) using the tilted ring method. Upon applying Eq.~\ref{eq:RC_fit} to fit the RC at this radius, the fitted RC (the orange solid line) clearly deviates from the truth (the black dashed line). We vary the orientation of the elliptical orbits to obtain the relation between the amplitude of the deviation and $\Delta \phi$, and we test different axial ratios of the elliptical orbits, as shown in Figure~\ref{fig:explanation}(c). The deviation increases with a higher axial ratio, especially when $\Delta \phi$ approaches zero. Indeed, the axial ratio of the nuclear ring in our simulations is approximately $1.5$. The deviation in the nuclear ring region (Figure~\ref{fig:RC_deviation}) demonstrates a similar trend and amplitude ($\sim30\%$) to the green line with the axial ratio ($a/b=1.5$) in Figure~\ref{fig:explanation}(c). 

In conclusion, the {\aell} model can reproduce similar trends between the deviation of the RC and $\Delta \phi$ in the nuclear ring region, but it fails to explain the asymmetry of the RC deviation with respect to $\Delta \phi=0\degree$ in Figure~\ref{fig:RC_deviation}. The limitations of the {\aell} model are evident; it is unable to capture the elliptical streamlines with varying orientations and axial ratios in the intermediate and bar regions. This limitation will be addressed in the subsequent model.

\subsection{{\Misaellt} Model}\label{subsec:misalign_ellipse}

To account for complex gas flows, we propose a {\misaell} model, which employs elliptical orbits with varying ellipticities ($\epsilon$) and orientations. This orbital arrangement is inspired by the kinematic density wave model proposed by \citet{1973PASA....2..174K}.

\begin{figure}
    \centering
    \includegraphics[width=1.0\columnwidth]{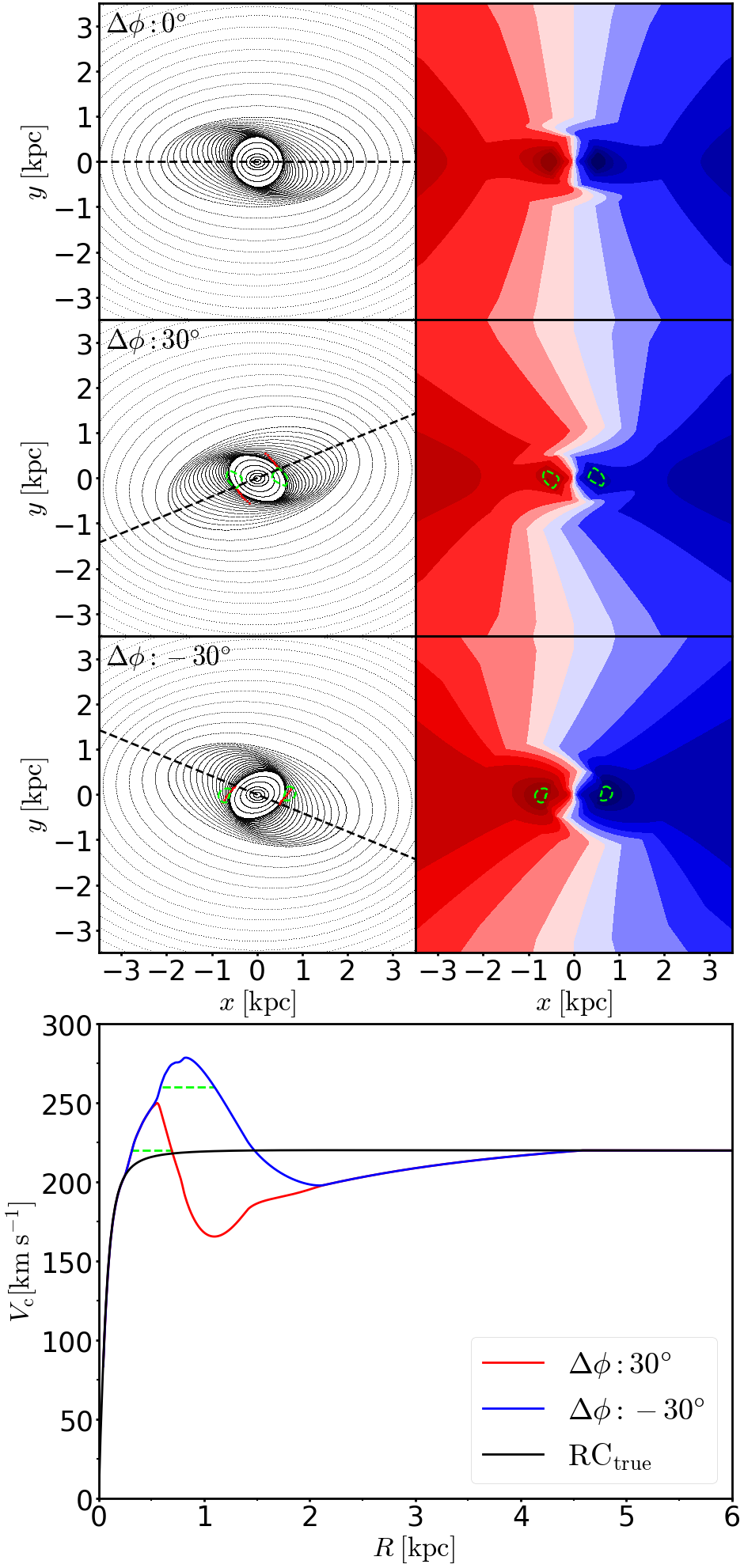}
    \caption{In the upper panel, the first column displays the distribution of orbits in our {\misaell} model with $\Delta \phi= 0\degree, \pm\; 30\degree$, projected at an inclination of $45\degree$. The black dashed lines indicate the orientation of the bar. The second column presents the contours of the $V_{\mathrm{los}}$ map. The green dashed lines represent velocity contours at ($\Delta \phi=30\degree:\pm\; 220\kms \times \sin{i}$, $\Delta \phi=-30\degree:\pm\; 260\kms \times \sin{i}$). The pericenters of the elliptical orbits are marked by red dots. The line of nodes is along the $x$-axis. The bottom panel presents a comparison of RCs for $\Delta\phi=\pm\;30\degree$. The black line represents $\trueRC$, while the red and blue lines represent $\tiltedRC$ when $\Delta\phi=\pm\;30\degree$. Two green dashed horizontal lines correspond to $V= 220\kms$ and $260\kms$, respectively.}
    \label{fig:Reason_pm_PA}
\end{figure}

The upper-left column in Figure~\ref{fig:Reason_pm_PA} illustrates this model. This panel shows the variation of elliptical orbits across different regions. Within the central region ($R\leq 0.25\kpc$), the orbits are circular. As we move from $0.25\kpc$ to the end of the nuclear ring, the ellipticity ($\epsilon$) of the orbits increases from zero to the ellipticity of the nuclear ring ($\epsilon_{\mathrm{ring}}\approx0.29$). The ellipticity $\epsilon_{\mathrm{ring}}$ is measured directly from the surface density of the simulation. Between the end of the nuclear ring and the bar end the ellipticity of the orbits increases from $\epsilon_{\mathrm{ring}}$ to the ellipticity of the bar, $\epsilon_{\mathrm{bar}}$($\approx 0.37$). Meanwhile, their orientations shift from perpendicular to parallel with respect to the bar. Outside the bar the orientations of the orbits remain unchanged, but their $\epsilon$ decreases from $\epsilon_{\mathrm{bar}}$ to zero. Each orbit is assigned a constant angular momentum based on the average radius and $\trueRC$ (the same as the {\aell} model).

We generate $V_{\rm los}$ maps using this {\misaell} model as shown in the upper-right column of Figure~\ref{fig:Reason_pm_PA}. For each $\Delta \phi$, we derive the RCs represented by the green dashed lines in columns 3 and 6 of Figure~\ref{fig:Sormani_Ferrers_RC}. Remarkably, this model can reproduce the bar-induced {\dip} features and capture the overall trend for $\left|\Delta \phi\right|\lesssim40\degree$ which is consistent with the results from the tilted-ring method. However, beyond this range of $\Delta \phi$, the model cannot create the high-velocity peaks observed in $\tiltedRC$. This is because our velocity assignment strategy is based on constant angular momentum. For a larger $\Delta \phi$, $V_\mathrm{los}$ has a large contribution from the radial inflows due to shocks, which is not included in the {\misaell} model.

A significant advantage of this model is its capacity to explain the difference in positive and negative $\Delta \phi$ cases. The second and third rows of Figure~\ref{fig:Reason_pm_PA} illustrate the orbit distribution and the $V_{\rm los}$ map when $\Delta \phi=\pm\;30\degree$. 
In a pure circular orbit model a symmetric ``spider'' pattern appears in the $V_{\rm los}$ map. At each radius the velocity adheres to a cosine or sine function of the azimuthal angle ($\phi$). The amplitude of this function corresponds to the RC, with its minimum and maximum values only appearing at the disk major axis. However, our {\misaell} model introduces twists and deviations from the spider pattern as the orientation of the orbits changes from perpendicular in the inner region to parallel with respect to the bar in the outer region. For instance, the areas enclosed by green dashed contours with $180\degree$ rotational symmetry in Figure~\ref{fig:Reason_pm_PA} comprise points where $V_{\rm los}$ exceeds $220\kms$ ($\Delta\phi=30\degree$) and $260\kms$($\Delta\phi=-30\degree$). These areas correspond to a peak in the RC higher than $220\kms$ ($\trueRC$). In the case of $\Delta\phi=-30\degree$, the green areas are closer to the pericenters (the red dots in Figure~\ref{fig:Reason_pm_PA}) of their respective orbits and further from the center compared to the scenario where $\Delta\phi=30\degree$. This results in a higher and outer peak in the RC, as the absolute value of $V_{\rm los}$ in these areas is higher than that in the positive $\Delta\phi$ case and significantly larger than that predicted by $\trueRC$.

To make our model more similar to the observation where the $V_{\rm los}$ and the density map are available but $\trueRC$ is unknown, we have modified our approach to assigning velocities to each elliptical orbit. Instead of relying on the $\trueRC$ we allocate velocities to each elliptical orbit directly from simulations. Furthermore, we incorporate additional constraints from the density image, including the positions of shocks and the morphology of the bar. For the derivation of the RC we fit the $V_\mathrm{los}$ for each orbit using an equation derived from an elliptical orbit with constant angular momentum:
\begin{equation}
\begin{array}{c}
    V_\mathrm{los}(\phi)= -L_\mathrm{z}(\frac{a}{b}\cos\phi \cos\phi_\mathrm{ell} + \frac{b}{a}\sin\phi\sin \phi_\mathrm{ell})f(\phi)\sin{i}, \\
    f(\phi) = \left(\frac{a^2\tan^2\phi+b^2}{a^4\tan^2\phi+b^4}\right)^{1/2}.
\end{array}
\label{eq:mell}
\end{equation}
Here, $L_{z},\ a,\ b,\ \phi_\mathrm{ell},$ and $\phi$ represent the angular momentum, the lengths of the major and minor axes, the position angle related to the disk major axis of the orbit, and the azimuthal angle of each point on the orbit, respectively. Additionally, we calculate the average radius ($\bar{R}$) for each orbit and define the circular velocity at $\bar{R}$ as $L_{z}/\bar{R}$. Detailed derivations can be found in Appendix~\ref{Ap:A}. $L_{z}(\bar{R})$ is the unknown parameter that we aim to determine by fitting the $V_\mathrm{los}$ map of the mock galaxy with Eq.~\ref{eq:mell}.

The findings from this fitting procedure are presented in Appendix~\ref{Ap:C}. Compared to the mock data the {\misaell} model only covers the nuclear ring region and the position of shocks, but it ignores the spirals beyond the bar. A noticeable elliptical discontinuity is observed in the $V_\mathrm{los}$ map generated by our {\misaell} model but is absent in the mock data. The RC from $L_z$ in the fitting of Eq.~\ref{eq:mell} still deviates significantly from $\trueRC$. These outcomes indicate that the {\misaell} model is inadequate despite our efforts to match observations. On the other hand, even though the gas streamlines resemble nearly elliptical shapes, the assumption of angular momentum conservation may not hold true. This discrepancy is expected because strong shocks occur in the bar region which violates the assumption. We attempt to integrate the conservation of the second integral of motion ($I_2$) or its proxy (e.g., \citealt{2021ApJ...913L..22Q}) in place of $L_z$; however, the application of $I_2$ to the elliptical streamlines in our model appears to be challenging.

In conclusion, the {\misaell} model, despite its limitations, can reasonably reproduce the bar-induced {\dip} features in $\tiltedRC$. This model suggests that the bar-induced {\dip} feature results from a projection effect of the elliptical streamlines. Specifically, the oppositely deviating trends of the RC in the bar and nuclear ring regions can be attributed to the different orientations of the gas streamlines. The contrast between these two regions contributes to the emergence of the bar-induced {\dip} feature in the RC. This model elucidates the significance of $\Delta \phi$ as a dominating factor governing the deviation of the RCs. Furthermore, its capacity to replicate the differences in positive and negative $\Delta \phi$ cases offers insights into the complex dynamics of gas flows in galaxies. Nevertheless, it remains challenging to quantitatively recover $\trueRC$ using this model.

\subsection{Correction for $\tiltedRC$}
\label{subsec:RCcorrection}
Given the limitations of the {\misaell} model in recovering the $\trueRC$, we propose a simple correction method inspired by \cite{2015A&A...578A..14C}. We assume that the relationship between $\Delta \phi$ and the deviations of RC, as empirically determined by our simulations in Figure~\ref{fig:RC_deviation}, is applicable to all galaxies. 

To implement this method, it is essential to normalize the relation for each individual galaxy. Specifically:

\begin{itemize}
    \item[-]We define the value at the outermost point of the RC as the flat component ($V_\mathrm{flat}$) for each galaxy. Compared to the quadrupole model, the difference in $V_\mathrm{flat}$ across galaxies arises from differences in mass. The deviation of the RC in Figure~\ref{fig:RC_deviation} is dimensionless; thus, we integrate this deviation with $V_\mathrm{flat}$ to make it applicable to each galaxy. 
    \item[-]For each galaxy and simulation, we align the minimum of the bar-induced {\dip} feature, splitting the region inside the bar into two: $[0,R_\mathrm{min})$ and $[R_\mathrm{min},R_\mathrm{bar}]$. 
    \item[-]We adjust the empirical deviation profile through stretching or compressing to derive a radial profile of the correction factor for each galaxy. Subsequently, this radial profile is applied to obtain the corrected RC for each galaxy, as represented by the red dots in Figure~\ref{fig:dip_galaxies}. 
\end{itemize}

These three steps can be expressed in the following equations:
\begin{equation}
    V_\mathrm{c,corrected}(R,\Delta\phi)=V_\mathrm{c,titled}(R)-G(R,\Delta \phi)V_\mathrm{flat}, 
\end{equation}
\begin{equation}
\begin{array}{c}
    G(R,\Delta\phi)= \left\{\begin{array}{cc}
         g(\frac{R_\mathrm{min,0}}{R_\mathrm{min}}\cdot R ,\Delta\phi),&[0,R_\mathrm{min})\\      g(k(R),\Delta\phi),  &[R_\mathrm{min}, R_\mathrm{bar}] 
    \end{array}\right.\\
\end{array}
\end{equation}

\begin{equation}
    k(R) = R_\mathrm{min,0}+\frac{R_\mathrm{bar,0}-R_\mathrm{min,0}}{R_\mathrm{bar}-R_\mathrm{min}}(R-R_\mathrm{min}),
\end{equation}
\begin{equation}
    g(R, \Delta \phi) = \frac{V_\mathrm{c,tilted,0}(R,\Delta \phi)-V_\mathrm{c,true,0}(R)}{V_\mathrm{flat,0}},
\end{equation}
where $R$, $R_\mathrm{min}$, $R_\mathrm{bar}$, and $\Delta \phi$ represent the galactocentric cylindrical radius, the radius at the minimum of the {\dip} feature, the radius at the bar length, and the position angle difference between the major axes of the bar and the disk in the face-on view, respectively, for the target galaxy. The quantities with a subscript of $0$ are derived from the quadrupole model. $R_\mathrm{bar,0}$ is $2r_q=4\kpc$, and $V_\mathrm{flat,0}=220\kms$. $R_\mathrm{min,0}$ is sensitive to the choice of $\Delta \phi$ as shown in Figure~\ref{fig:Sormani_Ferrers_RC}. To derive $g(R, \Delta \phi)$, we first adjust the quadrupole model to the same $\Delta \phi$ as the target galaxy to construct the $V_\mathrm{los}$ map. We then obtain $V_\mathrm{c,tilted,0}$ under this $\Delta \phi$ using the tilted-ring method. The profile of $g(R, \Delta \phi)$ therefore depends both on $R$ and $\Delta \phi$. The profile of $g(R, \Delta \phi)$ for three different radial bins is shown in the bottom panel of Figure~\ref{fig:RC_deviation}.

To facilitate comparison with the results from \cite{2020ApJ...897..122L}, we perform the same fitting for the corrected RCs using a simple equation (\citealt{1997AJ....114.2402C,2020ApJ...897..122L}): 
\begin{equation}
    V_\mathrm{c}(R)=V_0\frac{2}{\pi}\mathrm{arctan}(R/r_\mathrm{t}).
    \label{eq:RCfit}
\end{equation}
This function describes a smooth rise in rotational velocity to a maximum $V_0$ at infinity.

A comparison of the fitted RCs from our correction (red dashed lines) and those from \cite{2020ApJ...897..122L} (blue dashed lines) is presented in Figure~\ref{fig:dip_galaxies}. For most barred galaxies in the figure, the fitting by \cite{2020ApJ...897..122L} ignores the correction of the bar-induced {\dip} features, resulting in an underestimation of rotation velocities in the bar region. Considering the systematic deviation of RC, our corrected RCs may be closer to $\trueRC$.

Nevertheless, this is only a first-order correction to the bar-induced non-circular motions in the rotation curve. Simulations with the quadrupole model cannot predict the deviation of the RC for all observed galaxies, as the non-circular motions in different galaxies can vary greatly due to their different bar properties. Furthermore, a massive bulge may also create a decline-rise feature, as noted in \S~\ref{sec:noncirobs}. A more accurate correction for the rotation curve may require detailed dynamical modeling of these galaxies, which will be presented in a follow-up study. 

\section{Summary}
\label{sec:summary}

This paper presents a study of the rotation curves of barred galaxies by combining data from the PHANGS-ALMA survey with hydrodynamic simulations. The principal findings are as follows:

1. The $\tiltedRC$ of a simulated barred galaxy deviates from its $\trueRC$ due to the non-circular gaseous motions induced by the bar. This effect depends sensitively on the angle difference between the major axes of the bar and the disk, denoted as $\Delta \phi$. For $\left|\Delta \phi\right|\lesssim 40\degree$ where the bar is roughly parallel to the disk, non-circular motions would generate a bar-induced {\dip} feature in the RC. Basically, the {\dip} feature is caused by the perpendicular orientation of the gas flows in the nuclear ring and the bar.  At low $\left|\Delta \phi\right|$, streamlines in the nuclear ring tend to enhance $V_\mathrm{los}$, while those in the bar tend to suppress $V_\mathrm{los}$.

2. The bar-induced {\dip} features identified in simulations are also commonly found in real barred galaxies in the PHANGS-ALMA survey.

3. A simple {\misaell} model can qualitatively explain the deviations and the general trend of $\tiltedRC$, but difficulties remain for fully correcting the non-circular motion effects to accurately reproduce $\trueRC$. We propose a simple correction method to estimate $\trueRC$ and apply it to PHANGS-ALMA data.

While we attempt to correct RCs for each galaxy in our sample, the complexity of this task is greater than anticipated due to various factors, including the strength, length, and pattern speed of the bar. Consequently, developing a single comprehensive model for all galaxies is impractical. Therefore, this paper aims to highlight the deviation pattern of the RC caused by bar-induced non-circular motions and to provide a straightforward explanation and correction method. A more effective strategy may involve constructing sophisticated mass models and performing simulations with these models to replicate the specific features observed in each individual galaxy. This detailed approach will be presented in our upcoming papers.

------------ Acknowledgements ------------ 

\section*{Acknowledgements} 
The authors would like to thank Jerry Sellwood and Kristine Spekkens for their insightful discussions on {\Diskfit}, and the valuable suggestions from Ortwin Gerhard. The research presented here is partially supported by the National Key R\&D Program of China under grant No. 2018YFA0404501; by the National Natural Science Foundation of China under grant Nos. 12103032, 12025302, 11773052, 11761131016 (NSFC-DFG); by the ``111'' Project of the Ministry of Education of China under grant No. B20019; and by the China Manned Space Project under grant No. CMS-CSST-2021-B03. J.S. also acknowledges support from a \textit{Newton Advanced Fellowship} awarded by the Royal Society and the Newton Fund. This work made use of the Gravity Supercomputer at the Department of Astronomy, Shanghai Jiao Tong University, and the facilities of the Center for High Performance Computing at Shanghai Astronomical Observatory.

\software{Athena++\citep{2020ApJS..249....4S},
          Scipy \citep{2020NatMe..17..261V},
          Numpy \citep{2020Natur.585..357H}, 
          Jupyter notebook \citep{2016ppap.book...87K},
          IRAF \citep{1986SPIE..627..733T,1993ASPC...52..173T}
          }
\clearpage

\appendix
\restartappendixnumbering
\section{Density Threshold Tests using mock Data }\label{Appendix:dens_threshold}
\begin{figure}[ht]
    \centering
    \includegraphics[width=0.90\columnwidth]{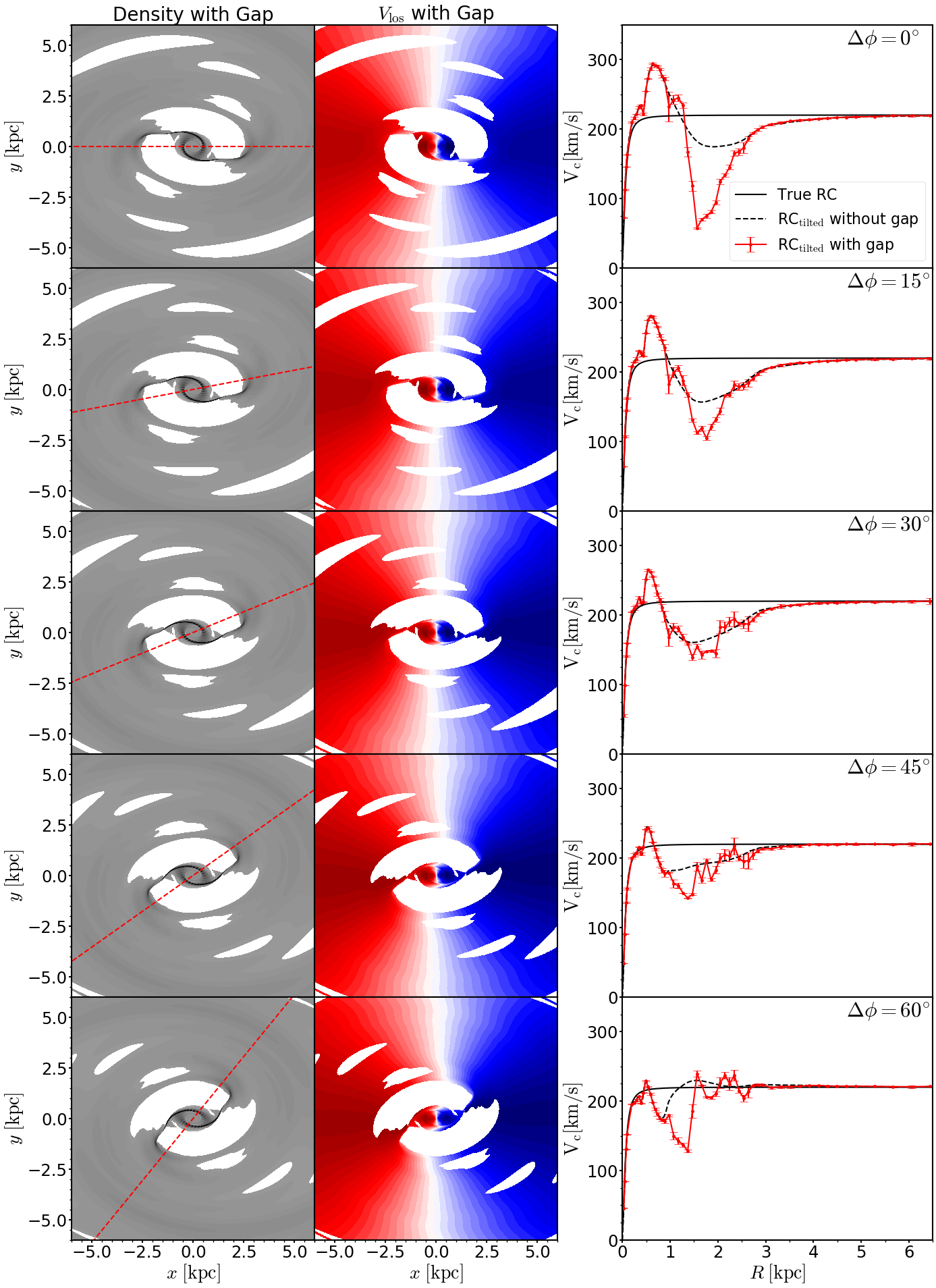}
    \caption{RCs from simulations with density threshold and different $\Delta \phi$ (from top to bottom, $\Delta \phi= 15\degree,\;30\degree,\;45\degree,\;60\degree$). The three columns show the projected surface density, $V_{\rm los}$ and RC comparisons, respectively. The red dashed lines in the first column represent the direction of the bar major axis. In Column 3 the black solid line, the red line with error bars and the black dashed line represent $\trueRC$, $\tiltedRC$ with and without a density threshold, respectively.
    }
    \label{fig:DenThres}
\end{figure}
To mimic the observational constraint in our simulations, we identify low-SN pixels using an empirical density threshold ($\Sigma_\mathrm{crit}=0.85\;\mathrm{M}_{\odot}\pc^{-2}$). Pixels with a surface density below $\Sigma_\mathrm{crit}$ are designated as {\tt NaN}, creating gaps in both the projected surface density and the $V_\mathrm{los}$ map, as shown in Figure~\ref{fig:DenThres}, Columns 1 and 2. The $\tiltedRC$ (red solid lines with error bars) is derived from a gap-containing $V_\mathrm{los}$ map and compared to the $\trueRC$ (black solid lines) and the gap-free $\tiltedRC$ (black dashed lines). The bar-induced {\dip} feature becomes more prominent with the density threshold.   
\section{Rotation Curves for galaxies without a {\DIP} feature}
\label{sec:appendixb}
\begin{figure*}[htb]
    \centering
    \includegraphics[width=1\textwidth]{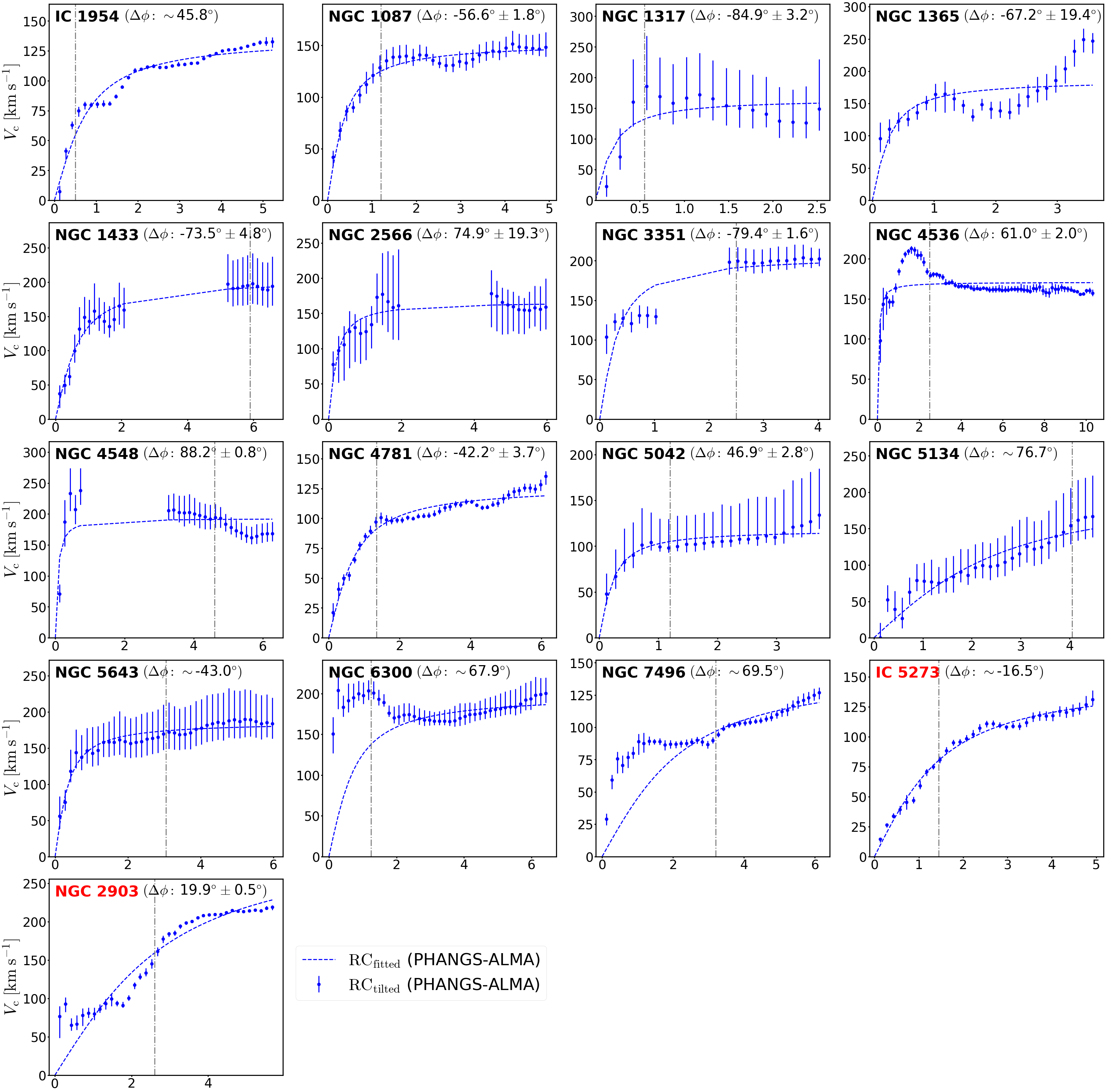}
    \caption{RCs for barred galaxies without the {\dip} feature. The blue dots with error bars represent the RC from PHANGS-ALMA. The blue dashed lines represent the best fit using Eq.~\ref{eq:RCfit}. The correction of RC is not applicable to these galaxies. The red-marked galaxies are the outliers that we discuss in the text. The grey horizontal dashed lines represent the bar length. The bar lengths of NGC~1365 and NGC~2566 are $10.2\kpc$ and $6.9\kpc$, respectively.}
    \label{fig:bar_galaxies_appb}
\end{figure*}

\begin{figure*}[htb]
    \centering
    \includegraphics[width=1.0\textwidth]{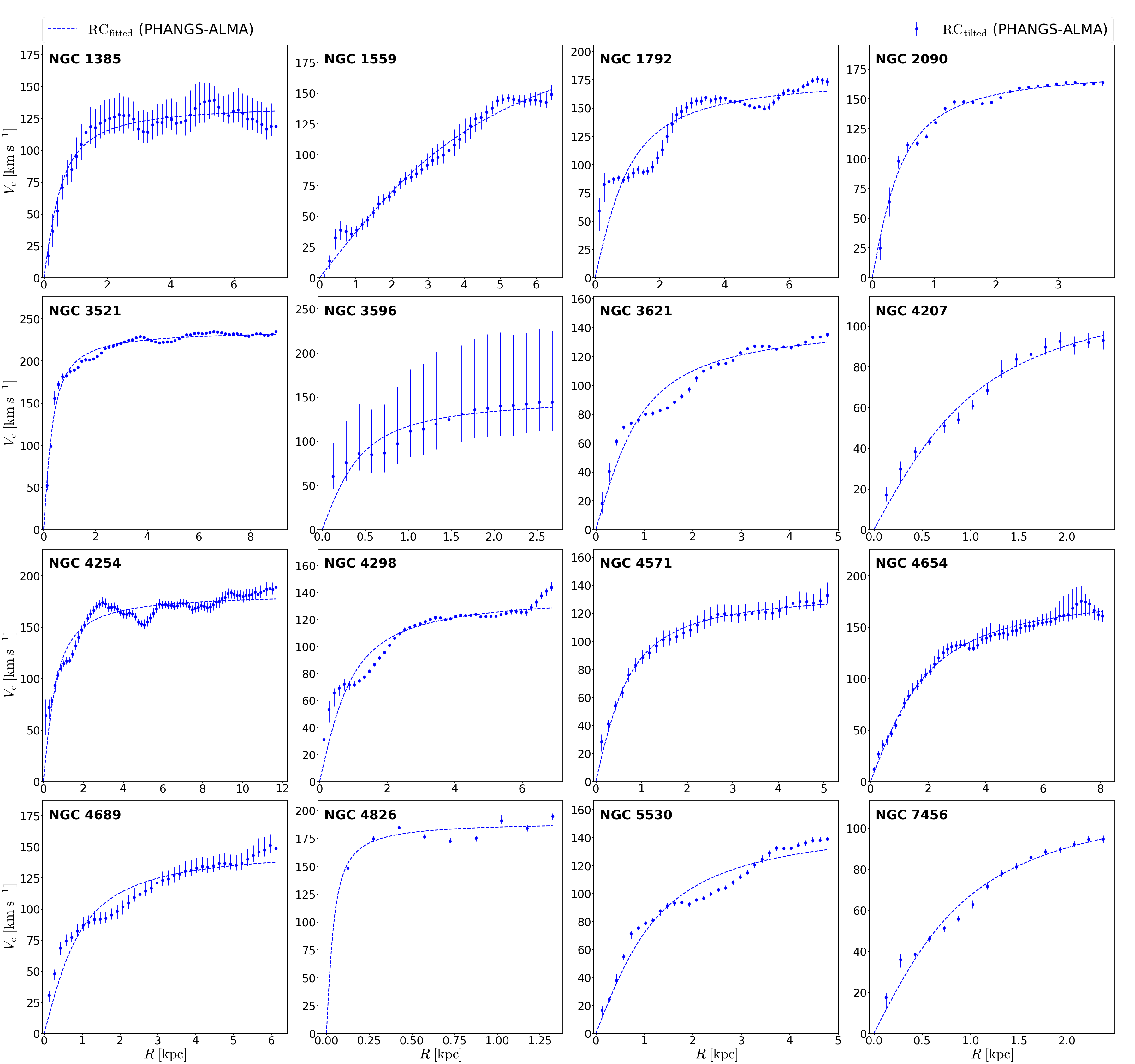}
    \caption{Same as Figure~\ref{fig:bar_galaxies_appb}, but for unbarred galaxies.}
    \label{fig:unbar_galaxies_appb}
\end{figure*}

We present rotation curves for barred (Figure~\ref{fig:bar_galaxies_appb}) and unbarred (Figure~\ref{fig:unbar_galaxies_appb}) galaxies without a {\dip} feature from the PHANGS-ALMA survey as supplementary material to support the discussions in \S~\ref{sec:noncirobs}.
\section{Analytic $V_{\rm \lowercase{los}}$ on an elliptical orbit}
\label{Ap:A}
The deviation starts from an elliptical orbit with its major axis parallel to the major axis of the projected disk ($\phi_\mathrm{ell}=0\degree$). 
\begin{equation}
\begin{array}{c}
    \text{Cartesian coordinate: }\frac{x^2}{a^2}+\frac{y^2}{b^2}=1\  (a>b) \\
    \text{Cylindrical coordinate: }R = \frac{ab}{\sqrt{a^2\sin^2{\phi}+b^2\cos^2{\phi}}}
\end{array}
\label{eq:ellporbit}
\end{equation}
The parameters $a$ and $b$ denote the major and minor axes of the elliptical orbit, respectively. The velocity of each point on the elliptical orbit is along the tangential direction, and the rotation is in a counterclockwise direction. The derivative of Eq.~\ref{eq:ellporbit} is combined with the assumption of angular momentum conservation to determine the components of the velocity in the $x$ and $y$ directions ($v_x$ and $v_y$).
\begin{equation}
\begin{array}{c}
    L_z=x\cdot v_y-y\cdot v_x= V_{\rm c}(\bar{R})\cdot{\bar{R}}=\text{const}\\
    \bar{R}=\frac{1}{2\pi}\int^{2\pi}_{0}Rd\phi=\frac{1}{2\pi}\int^{2\pi}_{0} \frac{abd\phi}{\sqrt{a^2\sin^2{\phi}+b^2\cos^2{\phi}}}  \\
    \frac{dy}{dx} = \frac{b^2}{a^2}\frac{x}{y}=\frac{b^2}{a^2\tan{\phi}},\  \text{Let }g=\arctan{\dot{y}}\\
    v_x =\frac{L_z}{R}\cos{g},v_y = \frac{L_z}{R}\sin{g}
\end{array}
\end{equation}
$\bar{R}$ represents the average radius of the elliptical orbit. The components $v_x$ and $v_y$ can be expressed as functions of $R$ and $\phi$:
\begin{equation}
\begin{array}{c}
    \tan{g}=\frac{dy}{dx} = \frac{b^2}{a^2\tan{\phi}}\\
    \cos{g}= \sqrt{\frac{1}{1+\tan^2{g}}}=\sqrt{\frac{a^4\tan^2{\phi}}{a^4\tan^2{\phi}+b^4}},\ \sin{g}= \sqrt{\frac{\tan^2{g}}{1+\tan^2{g}}} =\sqrt{\frac{b^4}{a^4\tan^2{\phi}+b^4}}, \\
    v_x = -L_z\frac{a}{b}\left(\frac{a^2\tan^2{\phi}+b^2}{a^4\tan^2{\phi}+b^4}\right)^{1/2}\sin{\phi},\ v_y = L_z\frac{b}{a}\left(\frac{a^2\tan^2{\phi}+b^2}{a^4\tan^2{\phi}+b^4}\right)^{1/2}\cos{\phi}.
\end{array}
\end{equation}

The subsequent procedure involves projecting the ellipse with an inclination $i$ to determine $V_{\rm los}$ and considering that the major axis of the orbit is not aligned with the disk major axis ($\phi_\mathrm{ell}\neq0\degree$). 
\begin{equation}
\begin{array}{c}
    V_{\rm{los}}=(v_x\sin{\phi_\mathrm{ell}}-v_y\cos{\phi_\mathrm{ell}})\sin{i}=  -L_\mathrm{z}(\frac{a}{b}\cos\phi \cos\phi_\mathrm{ell} + \frac{b}{a}\sin\phi\sin\phi_\mathrm{ell})\left(\frac{a^2\tan^2{\phi}+b^2}{a^4\tan^2{\phi}+b^4}\right)^{1/2}\sin{i}
\end{array}
\end{equation}
\section{Additional Test with the {\MisaEllt} model }
\label{Ap:C}
\begin{figure}[htb]
    \centering
    \includegraphics[width=0.95\columnwidth]{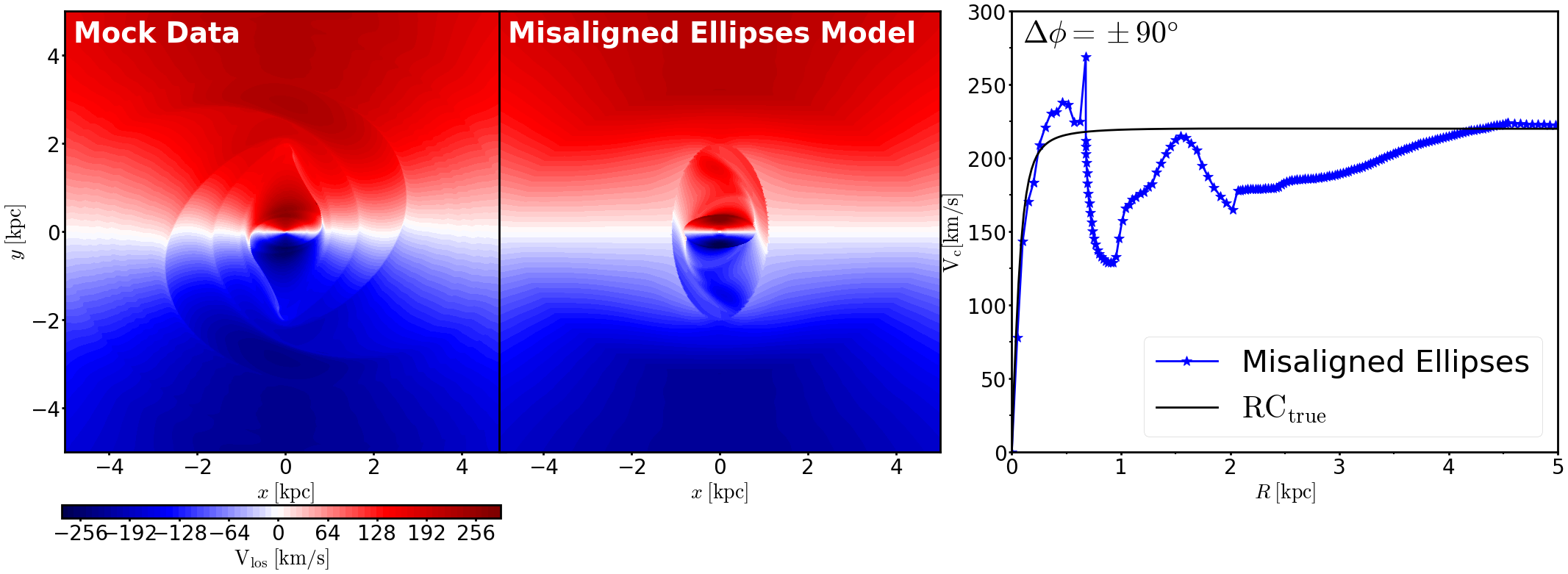}
    \caption{Results of the {\misaell} model with Eq.~\ref{eq:mell} when $\Delta\phi=\pm\;90\degree$. The left and middle panels compare the $V_\mathrm{los}$ map of mock data generated from simulation and our fitting with Eq.~\ref{eq:mell}. In the right panel the blue and black lines represent the RC from the {\misaell} model and $\trueRC$, respectively.
    }
    \label{fig:mell}
\end{figure}

As discussed in \S~\ref{subsec:misalign_ellipse}, we incorporate observational constraints (e.g., nuclear ring size, shock positions, kinematic map, etc.) into the {\misaell} model to quantitatively explain the deviation in the RC. Assuming constant angular momentum for each orbit, we derive it from the kinematic map using Eq.~\ref{eq:mell} (see Appendix~\ref{sec:appendixb}). Finally, based on the definition in \S~\ref{subsec:misalign_ellipse} the RC and the fitted kinematic map can be derived as shown in Figure~\ref{fig:mell}.   

\section{{\Diskfit} tests with different gas flow patterns}
\label{subsec:DiskFit}

The code {\Diskfit} \citep{2015arXiv150907120S} aims to quantitatively account for bar-induced non-circular motions with the bisymmetric flow using the following equations:
\begin{equation}
\begin{array}{c}
     V_\mathrm{model}(R,\phi) = V_\mathrm{sys}+\sin{i}\;[\bar{V}_\mathrm{t}(R)\cos{\phi}
     -V_\mathrm{2,t}(R)\cos{(2\theta_\mathrm{b})}\cos{\phi}-V_\mathrm{2,r}(R)\sin{(2\theta_\mathrm{b})}\sin{\phi}],\\\theta_b=\phi-\phi_b.
\end{array}
\end{equation}
The bisymmetric flow consists of two components: the tangential flow, $V_\mathrm{2,t}(R)$, and the radial flow, $V_\mathrm{2,r}(R)$. The model regards the average tangential velocity (or mean streaming speed), $\bar{V}_\mathrm{t}(R)$, as a proxy for the circular velocity, $V_\mathrm{c}(R)$. Additionally, {\Diskfit} has the capability to disable the bisymmetric flow, reverting to the conventional tilted-ring method. More details about this method are in \citet{2015arXiv150907120S}.

\begin{figure}[htb]
    \centering
    \includegraphics[width=0.95\columnwidth]{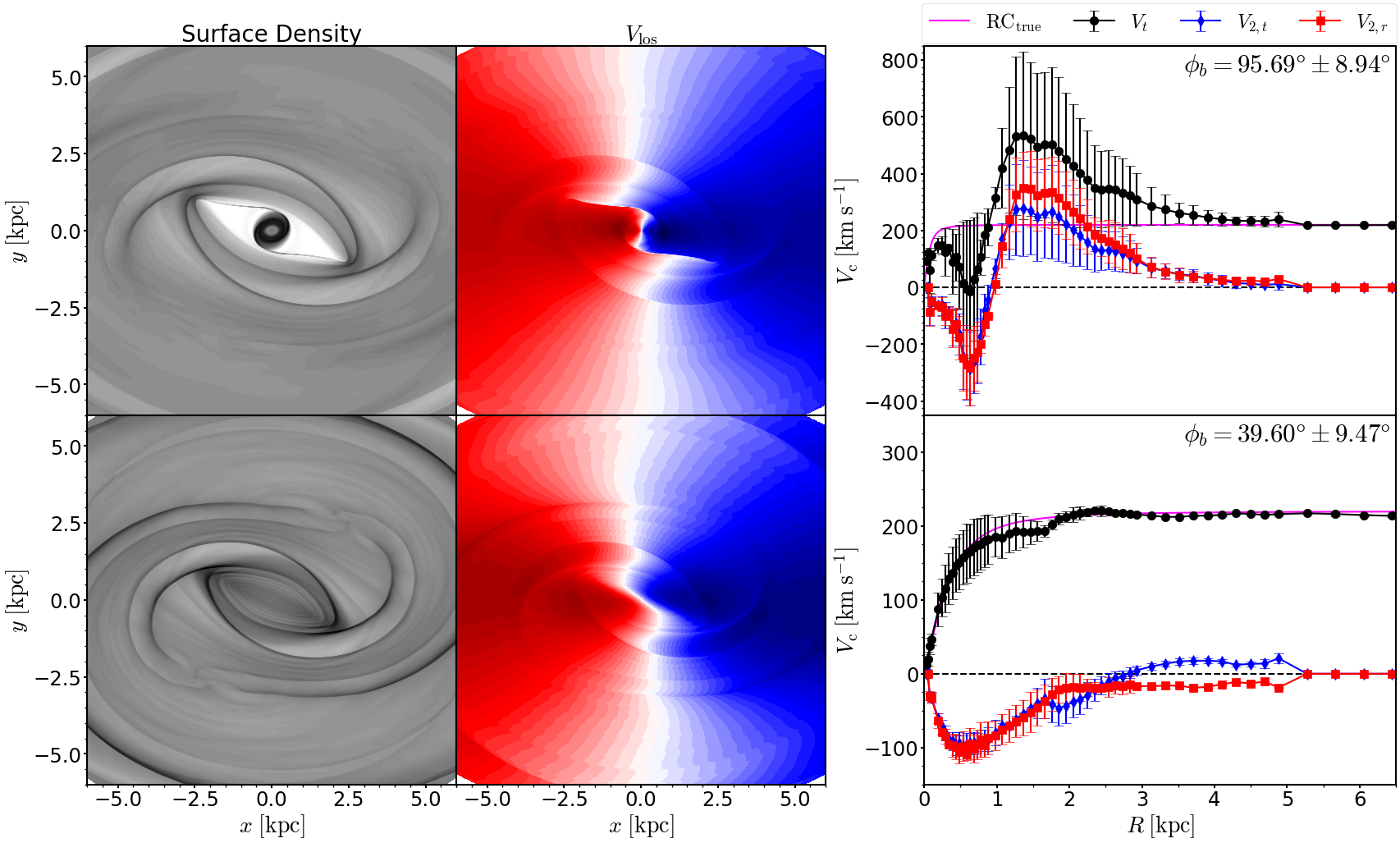}
    \caption{{\Diskfit} results for models with twisted and untwisted gas flow patterns at $\Delta \phi=-30\degree$. The first row shows the quadrupole model as presented in \S~\ref{sec:noncirsim}. The second row is the model with an untwisted flow pattern (without a nuclear ring). Column 1 and 2 show the projected surface density and the line-of-sight velocity for the two models, respectively. Column 3 compares $\trueRC$ with three components derived from the bisymmetric model. The magenta and black solid lines represent $\trueRC$ and the RC from the bisymmetric model of {\Diskfit}. The blue and red lines with error bars correspond to the tangential and radial components of the bisymmetric model, respectively. 
    }
    \label{fig:DiskFitPA0}
\end{figure}

The gas flow pattern in our simulation differs from that reported by \citet{2015MNRAS.454.3743R,2016A&A...594A..86R}. The central mass in their simulation seems not sufficiently concentrated to support the formation of a nuclear ring and the resultant twisted streamlines from the nuclear ring to the bar end. The absence of a nuclear ring results in an untwisted flow pattern within the bar, i.e., the orientation of the elliptical streamlines remains consistent and does not change from perpendicular to parallel to the bar major axis. We successfully generated a model with a similar untwisted pattern by increasing the $R_e$ parameter of the quadrupole model to $0.5\kpc$ (the second row of Figure~\ref{fig:DiskFitPA0}), with an intermediate $\Delta\phi=-30\degree$. The bisymmetric model is effective for the untwisted case, yielding an accurate RC close to $\trueRC$. However, the comparison suggests that a fixed phase angle in the bisymmetric model may not be appropriate for the twisted flow pattern with a nuclear ring, as commonly observed in both our simulation and real galaxies. 

\section{Comparison of Fitting Parameters}
\label{Ap:E}
We present the values of $V_0$ and $r_\mathrm{t}$, which have been derived from our corrected RC and \citet{2020ApJ...897..122L}.
\begin{table}[htp]
\begin{tabular}{|c||c|c||c|c|}
    \hline
    Galaxy &  $V_0\;(\kms)$ & $r_\mathrm{t}\;(\kpc)$ & $V_\mathrm{0,Lang}\;(\kms)$ &  $r_\mathrm{t,Lang}\;(\kpc)$ \\
    \hline
    NGC~1097 & 257.3 & 0.10 & 328.4 & 2.23\\
    NGC~1300 & 204.1 & 0.41 & 183.3 & 0.16\\
    NGC~3507 & 190.0 & 0.67 & 180.9 & 0.44\\
    NGC~3627 & 203.2 & 0.28 & 202.1 & 0.45\\
    NGC~4303 & 187.9 & 0.50 & 178.2 & 0.28\\
    NGC~4535 & 164.1 & 0.25 & 229.6 & 2.67\\
    NGC~4579 & 330.9 & 1.90 & 314.0 & 1.50\\
    NGC~4941 & 170.3 & 0.23 & 232.0 & 1.62\\
    NGC~5248 & 173.4 & 0.17 & 196.5 & 1.02\\
    \hline
\end{tabular}
    \caption{Parameters for fitting the rotation curve using Eq.~\ref{eq:RCfit}. Columns 2 and 3 are fitted using our corrected RC. Columns 4 and 5 present the value from \citet{2020ApJ...897..122L}.}
    \label{tab:RCfit}
\end{table}
\bibliography{sample631}{}

\begin{thebibliography}{}
\expandafter\ifx\csname natexlab\endcsname\relax\def\natexlab#1{#1}\fi
\providecommand{\url}[1]{\href{#1}{#1}}
\providecommand{\dodoi}[1]{doi:~\href{http://doi.org/#1}{\nolinkurl{#1}}}
\providecommand{\doeprint}[1]{\href{http://ascl.net/#1}{\nolinkurl{http://ascl.net/#1}}}
\providecommand{\doarXiv}[1]{\href{https://arxiv.org/abs/#1}{\nolinkurl{https://arxiv.org/abs/#1}}}

\bibitem[{{Athanassoula}(1992{\natexlab{a}})}]{1992MNRAS.259..345A}
{Athanassoula}, E. 1992{\natexlab{a}}, \mnras, 259, 345,
  \dodoi{10.1093/mnras/259.2.345}

\bibitem[{{Athanassoula}(1992{\natexlab{b}})}]{1992MNRAS.259..328A}
---. 1992{\natexlab{b}}, \mnras, 259, 328, \dodoi{10.1093/mnras/259.2.328}

\bibitem[{{Binney} {et~al.}(1991){Binney}, {Gerhard}, {Stark}, {Bally}, \&
  {Uchida}}]{1991MNRAS.252..210B}
{Binney}, J., {Gerhard}, O.~E., {Stark}, A.~A., {Bally}, J., \& {Uchida}, K.~I.
  1991, \mnras, 252, 210, \dodoi{10.1093/mnras/252.2.210}

\bibitem[{{Bosma}(1978)}]{1978PhDT.......195B}
{Bosma}, A. 1978, PhD thesis, University of Groningen, Netherlands

\bibitem[{{Chemin} {et~al.}(2015){Chemin}, {Renaud}, \&
  {Soubiran}}]{2015A&A...578A..14C}
{Chemin}, L., {Renaud}, F., \& {Soubiran}, C. 2015, \aap, 578, A14,
  \dodoi{10.1051/0004-6361/201526040}

\bibitem[{{Courteau}(1997)}]{1997AJ....114.2402C}
{Courteau}, S. 1997, \aj, 114, 2402, \dodoi{10.1086/118656}

\bibitem[{{Elmegreen}(1997)}]{1997RMxAC...6..165E}
{Elmegreen}, B.~G. 1997, in Revista Mexicana de Astronomia y Astrofisica
  Conference Series, Vol.~6, Revista Mexicana de Astronomia y Astrofisica
  Conference Series, ed. J.~{Franco}, R.~{Terlevich}, \& A.~{Serrano}, 165

\bibitem[{{Erwin}(2004)}]{2004A&A...415..941E}
{Erwin}, P. 2004, \aap, 415, 941, \dodoi{10.1051/0004-6361:20034408}

\bibitem[{{Grosb{\o}l} \& {Dottori}(2009)}]{2009A&A...499L..21G}
{Grosb{\o}l}, P., \& {Dottori}, H. 2009, \aap, 499, L21,
  \dodoi{10.1051/0004-6361/200911805}

\bibitem[{{Harris} {et~al.}(2020){Harris}, {Millman}, {van der Walt},
  {Gommers}, {Virtanen}, {Cournapeau}, {Wieser}, {Taylor}, {Berg}, {Smith},
  {Kern}, {Picus}, {Hoyer}, {van Kerkwijk}, {Brett}, {Haldane}, {del R{\'\i}o},
  {Wiebe}, {Peterson}, {G{\'e}rard-Marchant}, {Sheppard}, {Reddy}, {Weckesser},
  {Abbasi}, {Gohlke}, \& {Oliphant}}]{2020Natur.585..357H}
{Harris}, C.~R., {Millman}, K.~J., {van der Walt}, S.~J., {et~al.} 2020, \nat,
  585, 357, \dodoi{10.1038/s41586-020-2649-2}

\bibitem[{{Kalnajs}(1973)}]{1973PASA....2..174K}
{Kalnajs}, A.~J. 1973, \pasa, 2, 174, \dodoi{10.1017/S1323358000013461}

\bibitem[{{Kim} {et~al.}(2012{\natexlab{a}}){Kim}, {Seo}, \&
  {Kim}}]{2012ApJ...758...14K}
{Kim}, W.-T., {Seo}, W.-Y., \& {Kim}, Y. 2012{\natexlab{a}}, \apj, 758, 14,
  \dodoi{10.1088/0004-637X/758/1/14}

\bibitem[{{Kim} {et~al.}(2012{\natexlab{b}}){Kim}, {Seo}, {Stone}, {Yoon}, \&
  {Teuben}}]{2012ApJ...747...60K}
{Kim}, W.-T., {Seo}, W.-Y., {Stone}, J.~M., {Yoon}, D., \& {Teuben}, P.~J.
  2012{\natexlab{b}}, \apj, 747, 60, \dodoi{10.1088/0004-637X/747/1/60}

\bibitem[{{Kluyver} {et~al.}(2016){Kluyver}, {Ragan-Kelley}, {P{\'e}rez},
  {Granger}, {Bussonnier}, {Frederic}, {Kelley}, {Hamrick}, {Grout}, {Corlay},
  {Ivanov}, {Avila}, {Abdalla}, {Willing}, \& {Jupyter Development
  Team}}]{2016ppap.book...87K}
{Kluyver}, T., {Ragan-Kelley}, B., {P{\'e}rez}, F., {et~al.} 2016, in IOS
  Press, 87--90, \dodoi{10.3233/978-1-61499-649-1-87}

\bibitem[{{Lang} {et~al.}(2020){Lang}, {Meidt}, {Rosolowsky}, {Nofech},
  {Schinnerer}, {Leroy}, {Emsellem}, {Pessa}, {Glover}, {Groves}, {Hughes},
  {Kruijssen}, {Querejeta}, {Schruba}, {Bigiel}, {Blanc}, {Chevance},
  {Colombo}, {Faesi}, {Henshaw}, {Herrera}, {Liu}, {Pety}, {Puschnig}, {Saito},
  {Sun}, \& {Usero}}]{2020ApJ...897..122L}
{Lang}, P., {Meidt}, S.~E., {Rosolowsky}, E., {et~al.} 2020, \apj, 897, 122,
  \dodoi{10.3847/1538-4357/ab9953}

\bibitem[{{Leroy} {et~al.}(2021){Leroy}, {Schinnerer}, {Hughes}, {Rosolowsky},
  {Pety}, {Schruba}, {Usero}, {Blanc}, {Chevance}, {Emsellem}, {Faesi},
  {Herrera}, {Liu}, {Meidt}, {Querejeta}, {Saito}, {Sandstrom}, {Sun},
  {Williams}, {Anand}, {Barnes}, {Behrens}, {Belfiore}, {Benincasa},
  {Be{\v{s}}li{\'c}}, {Bigiel}, {Bolatto}, {den Brok}, {Cao}, {Chandar},
  {Chastenet}, {Chiang}, {Congiu}, {Dale}, {Deger}, {Eibensteiner}, {Egorov},
  {Garc{\'\i}a-Rodr{\'\i}guez}, {Glover}, {Grasha}, {Henshaw}, {Ho}, {Kepley},
  {Kim}, {Klessen}, {Kreckel}, {Koch}, {Kruijssen}, {Larson}, {Lee}, {Lopez},
  {Machado}, {Mayker}, {McElroy}, {Murphy}, {Ostriker}, {Pan}, {Pessa},
  {Puschnig}, {Razza}, {S{\'a}nchez-Bl{\'a}zquez}, {Santoro}, {Sardone},
  {Scheuermann}, {Sliwa}, {Sormani}, {Stuber}, {Thilker}, {Turner}, {Utomo},
  {Watkins}, \& {Whitmore}}]{2021ApJS..257...43L}
{Leroy}, A.~K., {Schinnerer}, E., {Hughes}, A., {et~al.} 2021, \apjs, 257, 43,
  \dodoi{10.3847/1538-4365/ac17f3}

\bibitem[{{Li} {et~al.}(2015){Li}, {Shen}, \& {Kim}}]{2015ApJ...806..150L}
{Li}, Z., {Shen}, J., \& {Kim}, W.-T. 2015, \apj, 806, 150,
  \dodoi{10.1088/0004-637X/806/2/150}

\bibitem[{{Maciejewski}(2004{\natexlab{a}})}]{2004MNRAS.354..883M}
{Maciejewski}, W. 2004{\natexlab{a}}, \mnras, 354, 883,
  \dodoi{10.1111/j.1365-2966.2004.08253.x}

\bibitem[{{Maciejewski}(2004{\natexlab{b}})}]{2004MNRAS.354..892M}
---. 2004{\natexlab{b}}, \mnras, 354, 892,
  \dodoi{10.1111/j.1365-2966.2004.08254.x}

\bibitem[{{Merrifield}(1992)}]{1992AJ....103.1552M}
{Merrifield}, M.~R. 1992, \aj, 103, 1552, \dodoi{10.1086/116168}

\bibitem[{{Miyamoto} \& {Nagai}(1975)}]{1975PASJ...27..533M}
{Miyamoto}, M., \& {Nagai}, R. 1975, \pasj, 27, 533

\bibitem[{{Oman} {et~al.}(2019){Oman}, {Marasco}, {Navarro}, {Frenk}, {Schaye},
  \& {Ben{\'\i}tez-Llambay}}]{2019MNRAS.482..821O}
{Oman}, K.~A., {Marasco}, A., {Navarro}, J.~F., {et~al.} 2019, \mnras, 482,
  821, \dodoi{10.1093/mnras/sty2687}

\bibitem[{{Pfenniger}(1984)}]{1984A&A...134..373P}
{Pfenniger}, D. 1984, \aap, 134, 373

\bibitem[{{Qin} \& {Shen}(2021)}]{2021ApJ...913L..22Q}
{Qin}, Y.-J., \& {Shen}, J. 2021, \apjl, 913, L22,
  \dodoi{10.3847/2041-8213/abfdb2}

\bibitem[{{Querejeta} {et~al.}(2021){Querejeta}, {Schinnerer}, {Meidt}, {Sun},
  {Leroy}, {Emsellem}, {Klessen}, {Mu{\~n}oz-Mateos}, {Salo}, {Laurikainen},
  {Be{\v{s}}li{\'c}}, {Blanc}, {Chevance}, {Dale}, {Eibensteiner}, {Faesi},
  {Garc{\'\i}a-Rodr{\'\i}guez}, {Glover}, {Grasha}, {Henshaw}, {Herrera},
  {Hughes}, {Kreckel}, {Kruijssen}, {Liu}, {Murphy}, {Pan}, {Pety}, {Razza},
  {Rosolowsky}, {Saito}, {Schruba}, {Usero}, {Watkins}, \&
  {Williams}}]{2021A&A...656A.133Q}
{Querejeta}, M., {Schinnerer}, E., {Meidt}, S., {et~al.} 2021, \aap, 656, A133,
  \dodoi{10.1051/0004-6361/202140695}

\bibitem[{{Randriamampandry} {et~al.}(2015){Randriamampandry}, {Combes},
  {Carignan}, \& {Deg}}]{2015MNRAS.454.3743R}
{Randriamampandry}, T.~H., {Combes}, F., {Carignan}, C., \& {Deg}, N. 2015,
  \mnras, 454, 3743, \dodoi{10.1093/mnras/stv2147}

\bibitem[{{Randriamampandry} {et~al.}(2016){Randriamampandry}, {Deg},
  {Carignan}, {Combes}, \& {Spekkens}}]{2016A&A...594A..86R}
{Randriamampandry}, T.~H., {Deg}, N., {Carignan}, C., {Combes}, F., \&
  {Spekkens}, K. 2016, \aap, 594, A86, \dodoi{10.1051/0004-6361/201629081}

\bibitem[{{Rhee} {et~al.}(2004){Rhee}, {Valenzuela}, {Klypin}, {Holtzman}, \&
  {Moorthy}}]{2004ApJ...617.1059R}
{Rhee}, G., {Valenzuela}, O., {Klypin}, A., {Holtzman}, J., \& {Moorthy}, B.
  2004, \apj, 617, 1059, \dodoi{10.1086/425565}

\bibitem[{{Rogstad} {et~al.}(1974){Rogstad}, {Lockhart}, \&
  {Wright}}]{1974ApJ...193..309R}
{Rogstad}, D.~H., {Lockhart}, I.~A., \& {Wright}, M.~C.~H. 1974, \apj, 193,
  309, \dodoi{10.1086/153164}

\bibitem[{{Rubin} {et~al.}(1982){Rubin}, {Ford}, {Thonnard}, \&
  {Burstein}}]{1982ApJ...261..439R}
{Rubin}, V.~C., {Ford}, W.~K., J., {Thonnard}, N., \& {Burstein}, D. 1982,
  \apj, 261, 439, \dodoi{10.1086/160355}

\bibitem[{{Salo} {et~al.}(2015){Salo}, {Laurikainen}, {Laine}, {Comer{\'o}n},
  {Gadotti}, {Buta}, {Sheth}, {Zaritsky}, {Ho}, {Knapen}, {Athanassoula},
  {Bosma}, {Laine}, {Cisternas}, {Kim}, {Mu{\~n}oz-Mateos}, {Regan}, {Hinz},
  {Gil de Paz}, {Menendez-Delmestre}, {Mizusawa}, {Erroz-Ferrer}, {Meidt}, \&
  {Querejeta}}]{2015ApJS..219....4S}
{Salo}, H., {Laurikainen}, E., {Laine}, J., {et~al.} 2015, \apjs, 219, 4,
  \dodoi{10.1088/0067-0049/219/1/4}

\bibitem[{{Sanders} \& {Tubbs}(1980)}]{1980ApJ...235..803S}
{Sanders}, R.~H., \& {Tubbs}, A.~D. 1980, \apj, 235, 803,
  \dodoi{10.1086/157683}

\bibitem[{{Schoenmakers} {et~al.}(1997){Schoenmakers}, {Franx}, \& {de
  Zeeuw}}]{1997MNRAS.292..349S}
{Schoenmakers}, R.~H.~M., {Franx}, M., \& {de Zeeuw}, P.~T. 1997, \mnras, 292,
  349, \dodoi{10.1093/mnras/292.2.349}

\bibitem[{{Sellwood} \& {Spekkens}(2015)}]{2015arXiv150907120S}
{Sellwood}, J.~A., \& {Spekkens}, K. 2015, arXiv e-prints, arXiv:1509.07120,
  \dodoi{10.48550/arXiv.1509.07120}

\bibitem[{{Sheth} {et~al.}(2010){Sheth}, {Regan}, {Hinz}, {Gil de Paz},
  {Men{\'e}ndez-Delmestre}, {Mu{\~n}oz-Mateos}, {Seibert}, {Kim},
  {Laurikainen}, {Salo}, {Gadotti}, {Laine}, {Mizusawa}, {Armus},
  {Athanassoula}, {Bosma}, {Buta}, {Capak}, {Jarrett}, {Elmegreen},
  {Elmegreen}, {Knapen}, {Koda}, {Helou}, {Ho}, {Madore}, {Masters},
  {Mobasher}, {Ogle}, {Peng}, {Schinnerer}, {Surace}, {Zaritsky},
  {Comer{\'o}n}, {de Swardt}, {Meidt}, {Kasliwal}, \&
  {Aravena}}]{2010PASP..122.1397S}
{Sheth}, K., {Regan}, M., {Hinz}, J.~L., {et~al.} 2010, \pasp, 122, 1397,
  \dodoi{10.1086/657638}

\bibitem[{{Silk}(1997)}]{1997ApJ...481..703S}
{Silk}, J. 1997, \apj, 481, 703, \dodoi{10.1086/304073}

\bibitem[{{Sofue}(2017)}]{2017PASJ...69R...1S}
{Sofue}, Y. 2017, \pasj, 69, R1, \dodoi{10.1093/pasj/psw103}

\bibitem[{{Sormani} {et~al.}(2015){Sormani}, {Binney}, \&
  {Magorrian}}]{2015MNRAS.454.1818S}
{Sormani}, M.~C., {Binney}, J., \& {Magorrian}, J. 2015, \mnras, 454, 1818,
  \dodoi{10.1093/mnras/stv2067}

\bibitem[{{Spekkens} \& {Sellwood}(2007)}]{2007ApJ...664..204S}
{Spekkens}, K., \& {Sellwood}, J.~A. 2007, \apj, 664, 204,
  \dodoi{10.1086/518471}

\bibitem[{{Stone} {et~al.}(2020){Stone}, {Tomida}, {White}, \&
  {Felker}}]{2020ApJS..249....4S}
{Stone}, J.~M., {Tomida}, K., {White}, C.~J., \& {Felker}, K.~G. 2020, \apjs,
  249, 4, \dodoi{10.3847/1538-4365/ab929b}

\bibitem[{{Sun} {et~al.}(2023){Sun}, {Leroy}, {Ostriker}, {Meidt},
  {Rosolowsky}, {Schinnerer}, {Wilson}, {Utomo}, {Belfiore}, {Blanc},
  {Emsellem}, {Faesi}, {Groves}, {Hughes}, {Koch}, {Kreckel}, {Liu}, {Pan},
  {Pety}, {Querejeta}, {Razza}, {Saito}, {Sardone}, {Usero}, {Williams},
  {Bigiel}, {Bolatto}, {Chevance}, {Dale}, {Gensior}, {Glover}, {Grasha},
  {Henshaw}, {Jim{\'e}nez-Donaire}, {Klessen}, {Kruijssen}, {Murphy},
  {Neumann}, {Teng}, \& {Thilker}}]{2023ApJ...945L..19S}
{Sun}, J., {Leroy}, A.~K., {Ostriker}, E.~C., {et~al.} 2023, \apjl, 945, L19,
  \dodoi{10.3847/2041-8213/acbd9c}

\bibitem[{{Takamiya} \& {Sofue}(2000)}]{2000ApJ...534..670T}
{Takamiya}, T., \& {Sofue}, Y. 2000, \apj, 534, 670, \dodoi{10.1086/308770}

\bibitem[{{Tody}(1986)}]{1986SPIE..627..733T}
{Tody}, D. 1986, in Society of Photo-Optical Instrumentation Engineers (SPIE)
  Conference Series, Vol. 627, Instrumentation in astronomy VI, ed. D.~L.
  {Crawford}, 733, \dodoi{10.1117/12.968154}

\bibitem[{{Tody}(1993)}]{1993ASPC...52..173T}
{Tody}, D. 1993, in Astronomical Society of the Pacific Conference Series,
  Vol.~52, Astronomical Data Analysis Software and Systems II, ed. R.~J.
  {Hanisch}, R.~J.~V. {Brissenden}, \& J.~{Barnes}, 173

\bibitem[{{van der Hulst} {et~al.}(1992){van der Hulst}, {Terlouw}, {Begeman},
  {Zwitser}, \& {Roelfsema}}]{1992ASPC...25..131V}
{van der Hulst}, J.~M., {Terlouw}, J.~P., {Begeman}, K.~G., {Zwitser}, W., \&
  {Roelfsema}, P.~R. 1992, in Astronomical Society of the Pacific Conference
  Series, Vol.~25, Astronomical Data Analysis Software and Systems I, ed. D.~M.
  {Worrall}, C.~{Biemesderfer}, \& J.~{Barnes}, 131

\bibitem[{{Virtanen} {et~al.}(2020){Virtanen}, {Gommers}, {Oliphant},
  {Haberland}, {Reddy}, {Cournapeau}, {Burovski}, {Peterson}, {Weckesser},
  {Bright}, {van der Walt}, {Brett}, {Wilson}, {Millman}, {Mayorov}, {Nelson},
  {Jones}, {Kern}, {Larson}, {Carey}, {Polat}, {Feng}, {Moore}, {VanderPlas},
  {Laxalde}, {Perktold}, {Cimrman}, {Henriksen}, {Quintero}, {Harris},
  {Archibald}, {Ribeiro}, {Pedregosa}, {van Mulbregt}, \& {SciPy 1. 0
  Contributors}}]{2020NatMe..17..261V}
{Virtanen}, P., {Gommers}, R., {Oliphant}, T.~E., {et~al.} 2020, Nature
  Methods, 17, 261, \dodoi{10.1038/s41592-019-0686-2}

\bibitem[{{Walter} {et~al.}(2008){Walter}, {Brinks}, {de Blok}, {Bigiel},
  {Kennicutt}, {Thornley}, \& {Leroy}}]{2008AJ....136.2563W}
{Walter}, F., {Brinks}, E., {de Blok}, W.~J.~G., {et~al.} 2008, \aj, 136, 2563,
  \dodoi{10.1088/0004-6256/136/6/2563}

\end{thebibliography}
\bibliographystyle{aasjournal}

\end{document}